\documentclass[amsmath,amssymb,aps,prd,11pt,tightenlines,superscriptaddress,nofootinbib,preprintnumbers,notitlepage]{revtex4-2}

\usepackage{xspace}
\usepackage{siunitx}
\usepackage{array}
\usepackage{tabularx}
\usepackage{xcolor}
\usepackage{ulem}
\usepackage{lineno}
\usepackage{amsmath,amssymb,amsthm,amsfonts}
\usepackage{graphicx,tabularx}
\usepackage{color}
\usepackage{multirow}
\usepackage{comment}
\usepackage{enumitem}
\usepackage{subfigure}
\usepackage{orcidlink}
\usepackage{cleveref}
\usepackage{slashed}
\usepackage{scalerel}
\usepackage{wrapfig}
\usepackage{cancel}
\usepackage{xcolor}
\usepackage{multirow}
\usepackage{setspace}

\newcommand{\PRE}[1]{{#1}} 

\newcommand{\be}{\begin{equation} \begin{aligned}}
\newcommand{\ee}{\end{aligned} \end{equation}}
\newcommand{\beqa}{\begin{eqnarray}}
\newcommand{\eeqa}{\end{eqnarray}}

\def\figureautorefname~#1\null{Fig.\,#1\null}
\def\tableautorefname~#1\null{Tab.\,#1\null}
\def\equationautorefname~#1\null{Eq.\,(#1)\null}

\crefname{section}{Sec.}{Secs.}
\crefname{figure}{Fig.}{Figs.}
\crefname{equation}{Eq.}{Eqs.}
\crefname{table}{Table}{Tables}
\crefname{appendix}{Appendix}{Appendices}

\makeatletter
\renewcommand{\p@subsection}{}
\renewcommand{\p@subsubsection}{}
\makeatother

\newcommand{\FASERnu}{FASER$\nu$\xspace}

\begin{document}


\title{Momentum Measurement of Charged Particles in FASER's Emulsion Detector at the LHC 
\bigskip \\  
FASER Collaboration
}

\begin{figure*}[h]
\vspace*{-0.6in}
\begin{flushleft}
\includegraphics[width=0.19\textwidth]{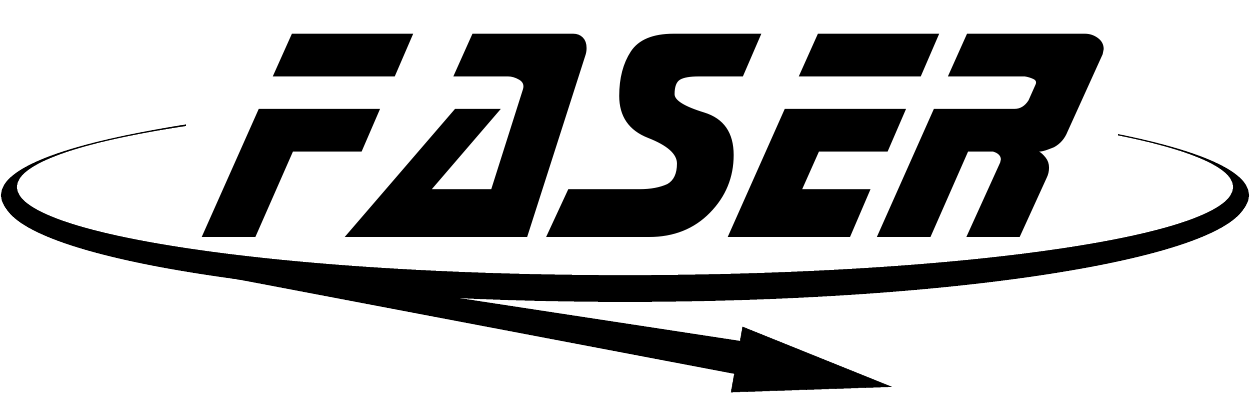}
\end{flushleft}
\end{figure*}

\author{Roshan Mammen Abraham\,\orcidlink{0000-0003-4678-3808}}
\affiliation{Department of Physics and Astronomy, University of California, Irvine, CA 92697-4575, USA}

\author{Xiaocong Ai\,\orcidlink{0000-0003-3856-2415}}
\affiliation{School of Physics, Zhengzhou University, Zhengzhou 450001, China}
  
\author{Saul Alonso Monsalve\,\orcidlink{0000-0002-9678-7121}}
\affiliation{Institute for Particle Physics, ETH Z\"urich, Z\"urich 8093, Switzerland}

\author{John Anders\,\orcidlink{0000-0002-1846-0262}}
\affiliation{University of Liverpool, Liverpool L69 3BX, United Kingdom}

\author{Emma Kate Anderson\,\orcidlink{0000-0002-0161-4560}}
\affiliation{CERN, CH-1211 Geneva 23, Switzerland}

\author{Claire Antel\,\orcidlink{0000-0001-9683-0890}}
\affiliation{D\'epartement de Physique Nucl\'eaire et Corpusculaire, University of Geneva, CH-1211 Geneva 4, Switzerland}

\author{Akitaka Ariga\,\orcidlink{0000-0002-6832-2466}}
\affiliation{Albert Einstein Center for Fundamental Physics, Laboratory for High Energy Physics, University of Bern, Sidlerstrasse 5, CH-3012 Bern, Switzerland}
\affiliation{Department of Physics, Chiba University, 1-33 Yayoi-cho Inage-ku, 263-8522 Chiba, Japan}

\author{Tomoko Ariga\,\orcidlink{0000-0001-9880-3562}}
\affiliation{Kyushu University, 744 Motooka, Nishi-ku, 819-0395 Fukuoka, Japan}

\author{Jeremy Atkinson\,\orcidlink{0009-0003-3287-2196}}
\affiliation{Albert Einstein Center for Fundamental Physics, Laboratory for High Energy Physics, University of Bern, Sidlerstrasse 5, CH-3012 Bern, Switzerland}

\author{Florian~U.~Bernlochner\,\orcidlink{0000-0001-8153-2719}}
\affiliation{Universit\"at Bonn, Regina-Pacis-Weg 3, D-53113 Bonn, Germany}

% ...

\begin{abstract}
\bigskip
We present a momentum measurement method based on multiple Coulomb scattering (MCS) in the \FASERnu emulsion detector.
The measurement of charged-particle momenta is essential for studying neutrino interactions in the TeV energy range at the FASER experiment.
This method exploits the sub-micron spatial resolution and long tracking length of the \FASERnu detector, enabling momentum determination from a few GeV up to a few TeV.
The performance was evaluated using Geant4-based Monte Carlo simulations and validated with muon test beam data in the momentum range 100-300 GeV.
As a first probe of the method for higher momentum muons, background muons recorded by the \FASERnu detector were examined, showing reconstructed momenta consistent with expectations from their angular spread.
\end{abstract}

\maketitle

\begin{center}
\copyright~2026 CERN for the benefit of the FASER Collaboration. Reproduction of this article or parts of it is allowed as specified in the CC-BY-4.0 license.  
\end{center}

\clearpage

\section{Introduction}

The ForwArd Search ExpeRiment (FASER)~\cite{Feng:2017uoz, FASER:2018ceo, FASER:2018bac, FASER:2022hcn} at CERN's Large Hadron Collider (LHC)~\cite{evans2008lhc} is designed to measure neutrino differential cross-sections in the TeV energy range with the neutrinos produced in the LHC’s proton-proton ($pp$) collisions using the dedicated \FASERnu emulsion detector~\cite{FASER:2019dxq, FASER:2020gpr}.
In 2021, the FASER Collaboration reported the first evidence of neutrino interaction candidates produced at the LHC~\cite{FASER:2021mtu}. 
In 2023, the first observation of collider muon neutrinos was achieved using FASER’s electronic detector components~\cite{FASER:2023zcr}. 
This observation was shortly after confirmed by the SND@LHC Collaboration~\cite{SND:2023du}.
With the electronic detector, FASER has also made the first measurement of the muon neutrino interaction cross section and flux as a function of energy~\cite{FASER:2024ref}.
In 2024, for the first time, FASER measured the $\nu_e$ and $\nu_\mu$ interaction cross sections using a $\SI{128}{kg}$ subset of the FASER$\nu$ detector after exposure to 9.5~fb$^{-1}$ of 13.6 TeV $pp$ collisions~\cite{FASER:2024hoe}. 
The \FASERnu detector consists of 730 layers of tungsten plates (\SI{1.1}{\milli\metre} thick) interleaved with emulsion films, each composed of a \SI{210}{\micro\metre} plastic base coated on both sides with \SI{65}{\micro\metre} emulsion layers.
The target volume has dimensions of approximately $\SI{25}{\centi\metre} \times \SI{30}{\centi\metre}\times \SI{108}{\centi\metre}$, corresponding to a target mass of 1.1 metric tons and a total depth of eight interaction lengths.

When a charged particle passes through the emulsion, silver bromide crystals are ionized, and after chemical development, clusters of silver grains are formed along its trajectory~\cite{ariga2020nuclear}.
Using the Hyper Track Selector (HTS), a high-speed automated track scanning system~\cite{Yoshimoto:2017ufm}, these grain sequences can be read out and reconstructed in three dimensions with very high precision, with a typical spatial resolution of 0.2 to 0.3 \SI{}{\micro\metre} in the transverse plane and a few \SI{}{\micro\metre} in the z direction.
Within a single emulsion layer, the trajectory of a particle is reconstructed as a micro-track, a sequence of aligned silver grains.
Two corresponding micro-tracks on each sides of the plastic base are connected to form a base-track, defined as the straight line linking the grains closest to the base surfaces.
Base-tracks are connected across multiple films to reconstruct the full three-dimensional particle trajectory. In the \FASERnu detector, the typical transverse position resolution is \SI{0.3}{\micro\metre}~\cite{FASERnu:Reconstruction}.

The momentum of charged particles can be measured by the effect of multiple Coulomb scattering (MCS) on the particle trajectory. 
Two main approaches are used for MCS-based momentum estimation. One is the angular method, which evaluates angular variations of the tracks, previously used in the DONuT~\cite{DONUT:Final} and OPERA~\cite{OPERA:2011aa} experiments. The other is the coordinate method, which analyzes positional deviations between successive plates and was used in the DONuT~\cite{DONUT:Mom} experiment.
The choice between the two methods depends on the required momentum range, the achievable spatial and angular resolutions, as well as the detector structure.
Although the angular method has the advantage of being insensitive to alignment errors, its applicability was limited to around 8 GeV in OPERA~\cite{OPERA:2011aa}.
In contrast, the coordinate method was validated up to 100 GeV through comparison with a spectrometer in DONuT~\cite{DONUT:Mom}, and it can in principle reach a few TeV, albeit requiring large statistics and very precise alignment.
Thanks to the excellent position resolution and long tracking length of the \FASERnu detector, the coordinate method can be effectively applied. By precisely measuring tiny deflections, reliable momentum determination of charged particles can be achieved over a wide range from a few GeV to a few TeV.
This range covers the typical momenta of primary muons and secondary hadrons produced in neutrino interactions observed by the FASER experiment, and the method has already been used for the kinematical selection of muon-neutrino interaction candidates requiring at least one track penetrating 100 plates with momentum above 200 GeV~\cite{FASER:2024hoe}.
In this paper, we evaluate the performance of the method using Monte Carlo (MC) simulations based on Geant4~\cite{Geant4:2003} from 10 GeV to 3 TeV and validate it with muon test-beam data in the range 100–300 GeV. 
As a first probe of the method at higher momenta, we further apply it to background muons recorded by the \FASERnu detector and examine the consistency of the reconstructed momenta with expectations.

This paper is organized as follows. \Cref{sec:MomentumMeasurementMethod} describes the momentum measurement method based on the coordinate method. In \Cref{sec:Evaluation}, we evaluate its performance using MC simulations. \Cref{sec:Testbeam} presents the muon test-beam exposure, the track reconstruction, and the validation of the measurement method using the obtained data. 
\Cref{sec:UsingF222} presents the application of this method to background muons recorded by the \FASERnu detector and discusses its performance in the TeV momentum range.
Finally, \Cref{sec:Conclusions} summarizes the conclusions.



\section{Momentum measurement method}
\label{sec:MomentumMeasurementMethod}

Momenta of charged particles are measured by the MCS effect on the trajectory measured in the FASER$\nu$ detector.
As a charged particle passes through matter, it undergoes numerous small-angle deflections due to elastic scatterings off nuclei.
The difference in angles before and after traversing the material peaks at zero and approximately follows a Gaussian distribution.
The RMS of the distribution of a singly charged particle is represented by the Highland formula~\cite{MCS:Bethe, MCS:Highland, PDG:2022yn}: 
\begin{equation}
\theta_{\mathrm{plane}}^{\mathrm{RMS}} = \frac{0.0136\ \mathrm{GeV}}{\beta pc}\sqrt{\frac{z}{X_0} }\left\{1 + 0.038\ln\left(\frac{z}{X_0 \beta^2} \right) \right\}.
\label{eq:thetaRMS}
\end{equation}
where $p$ is the momentum of the particle, $\beta c$ is velocity, $z$ is the thickness of the medium that the particle passed, and $X_0$ is the radiation length of the medium (\SI{3.504}{\milli\metre} in tungsten~\cite{PDG2024:AtomicNuclearProperties}). 

\begin{figure}[tbp]
    \centering
    \begin{minipage}{0.34\textwidth}
        \centering
        \includegraphics[width=\textwidth]{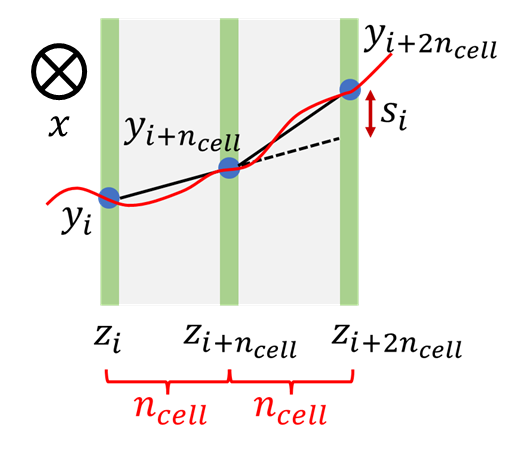}
    \end{minipage}
    \begin{minipage}{0.60\textwidth}
        \centering
        \includegraphics[width=\textwidth]{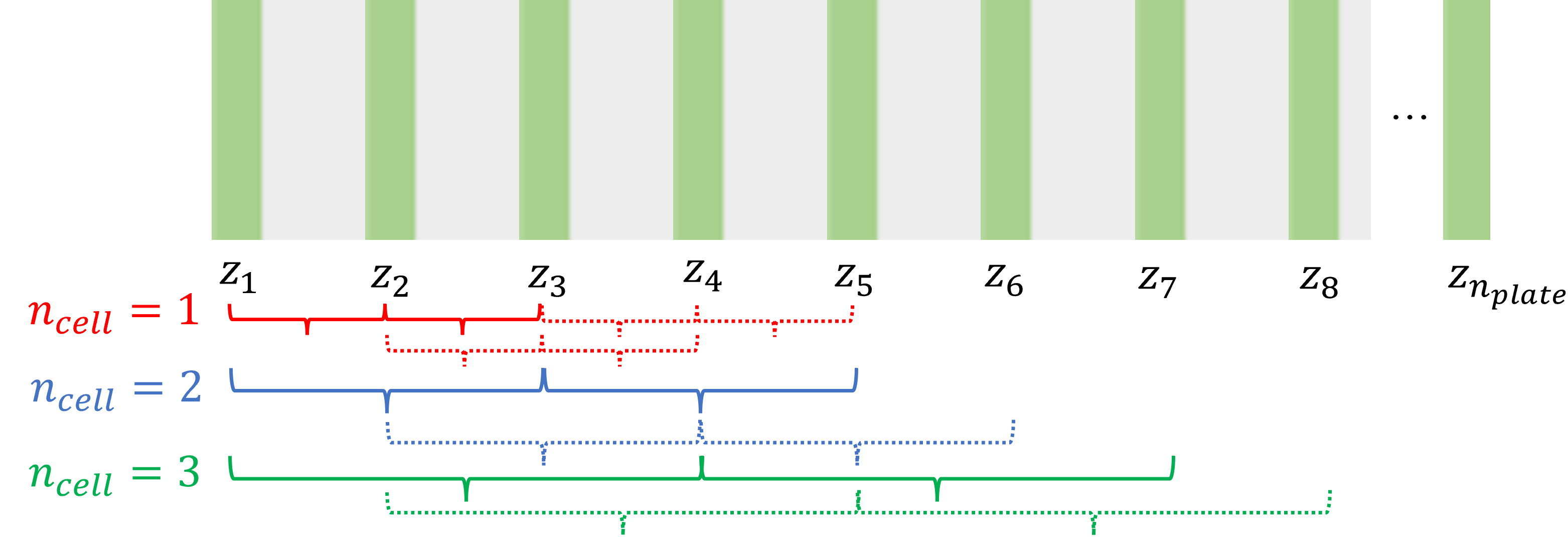}
    \end{minipage}
    \caption{Left: The second difference in $n_{\mathrm{cell}}$. Right: The schematic view of $n_{cell}$ = 1-3 and the shift of 1 plate when calculating the second difference (dotted line).}
    \label{fig:coordinate_scattering}
\end{figure}

The coordinate method utilizes position displacements, expressed as the second difference $s$, as shown on the left in \cref{fig:coordinate_scattering}. 
The position displacement for plate $i$, denoted as $s_i$, is given by the following equations:
\begin{align}
s_{i} = y_{i+2n_{\mathrm{cell}}} -\left\{ y_{i+n_{cell}} + \frac{y_{i+n_{\mathrm{cell}}}-y_{i}}{z_{i+n_{\mathrm{cell}}}-z_{i}} \left(z_{i+2n_{\mathrm{cell}}}-z_{i+n_{\mathrm{cell}}} \right) \right\}.
\label{eq:position_difference}
\end{align}
where $y_i$ represents the position coordinate, $z_i$ corresponds to the position along the beam axis, and $n_{\mathrm{cell}}$ is the cell length. 
The cell length is a unit of length used to measure the scattering over a given distance, as illustrated on the right in \cref{fig:coordinate_scattering}.
In this equation, the position displacement $s_i$ is measured using the positional information ($y$, $z$) in the three plates: plate $i$, plate $i+n_{\mathrm{cell}}$, and plate $i+2n_{\mathrm{cell}}$.
Two position coordinates, ($y_i$, $z_i$) and ($y_{i+n_{\mathrm{cell}}}$, $z_{i+n_{\mathrm{cell}}}$), are used to determine a line, and the third position coordinate, ($y_{i+2n_{\mathrm{cell}}}$, $z_{i+2n_{\mathrm{cell}}}$), is used to calculate the positional displacement due to MCS compared to the extrapolated line. 
The second difference $s$ can be calculated in each of $x$-$z$ and $y$-$z$ planes independently.  

\begin{figure}[h]
    \centering
    \includegraphics[width=0.40\linewidth]{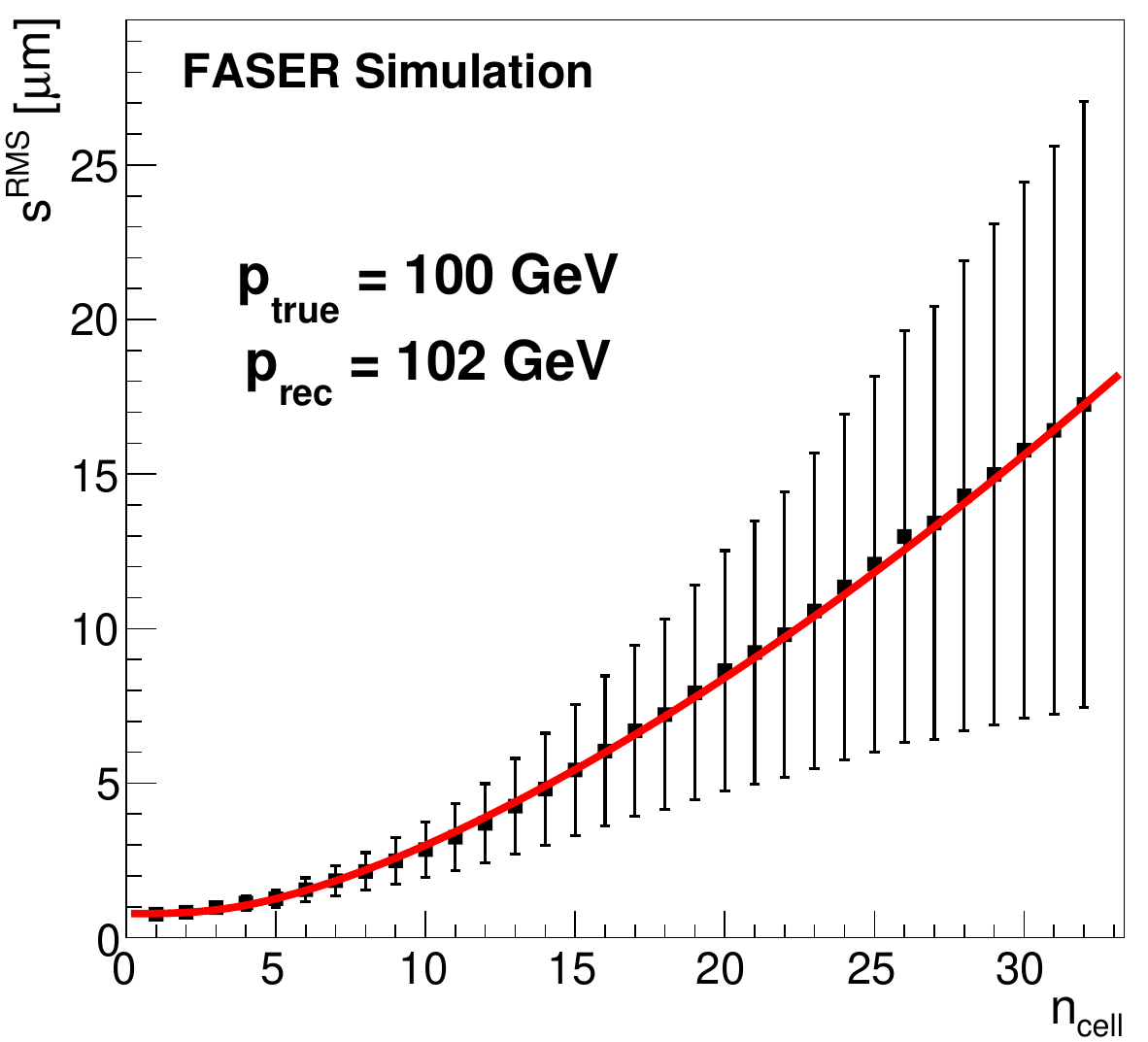}
    \includegraphics[width=0.40\linewidth]{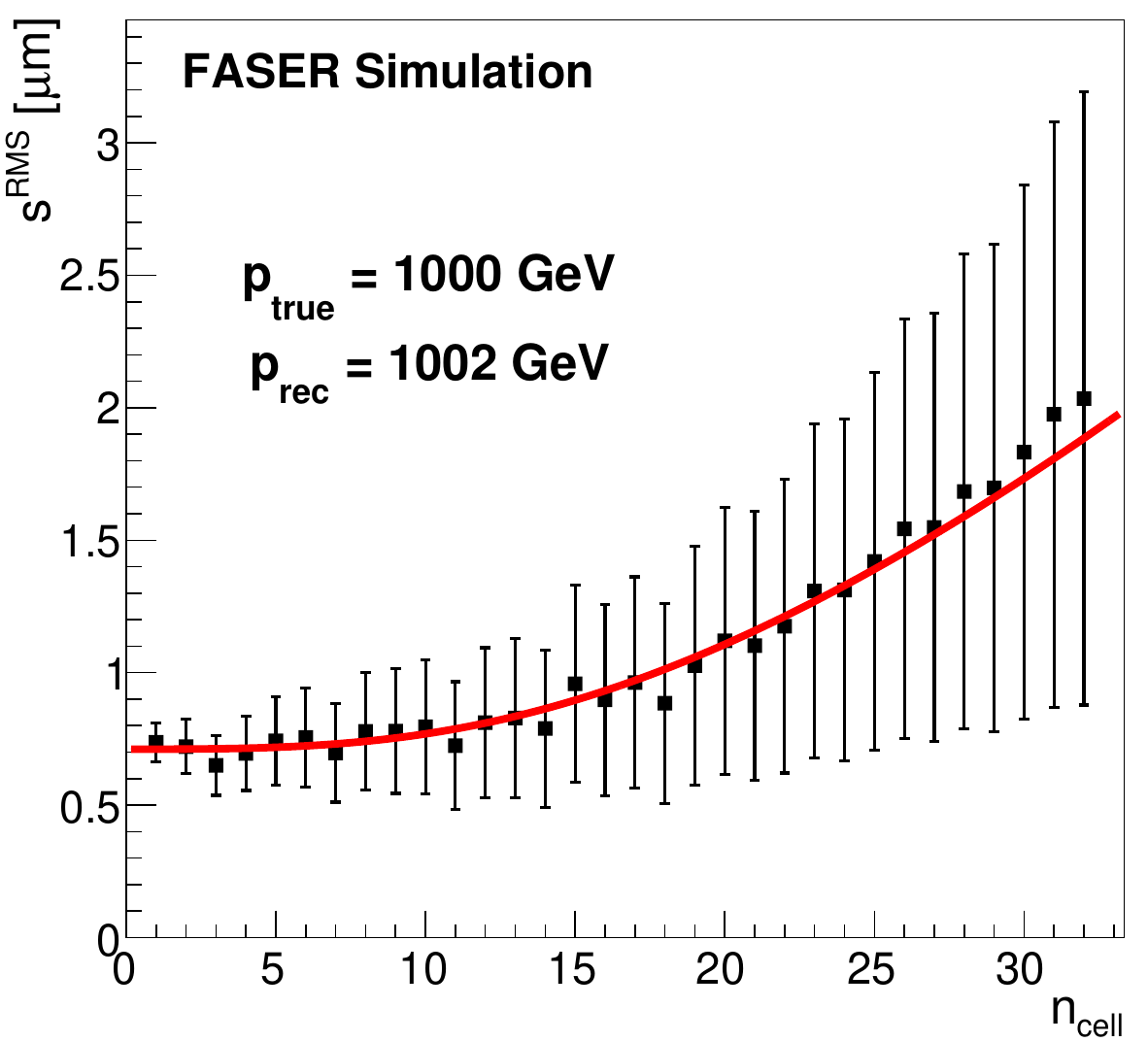}
    \caption{Examples of $s^{\mathrm{RMS}}$ measurements and momentum fitting for single tracks. Left: $p_{\mathrm{true}}$ = 100 GeV. Right: $p_{\mathrm{true}}$ = 1000 GeV. The black points are measured $s^{\mathrm{RMS}}$ (\SI{}{\micro\metre}) for each cell length and red line is the fit of the function in \cref{eq:srms}.
    The black bars represent the effective statistical weights. In the fitting procedure, $1/w(n_{\mathrm{cell}})$ is used for each data point.
    $p_{\mathrm{rec}}$ = 102 GeV for $p_{\mathrm{true}}$ = 100 GeV, and 1002 GeV for 1000 GeV.
    For this example, the tracks are generated with Geant4~\cite{Geant4:2003}, following the same setup as in \cref{sec:Evaluation}, are used. A total of 100 plates are considered, with a position smearing of $\sigma_{\mathrm{pos}}$ = \SI{0.3}{\micro\metre} and a base-track reconstruction efficiency of 90\%.}    
    \label{fig:PrecCoordFit}
\end{figure}

The second difference includes two contributions: the displacement due to MCS and the position resolution.
The contribution from MCS to the deviation of $s$, $s_{\mathrm{MCS}}^{\mathrm{RMS}}$ is expected to be
\begin{equation}
s_{\mathrm{MCS}}^{\mathrm{RMS}} \left( p, n_{\mathrm{cell}}\right)=  \sqrt{\frac{2}{3}}\cdot\frac{0.0136\ \mathrm{GeV}}{\beta pc}\cdot z_{n_{\mathrm{cell}}}\sqrt{\frac{z_{n_{\mathrm{cell}}}}{X_c}}\left\{1 + 0.038 \ln\left(\frac{z_{n_{\mathrm{cell}}}}{X_c\beta^2} \right) \right\}.
\label{eq:splanerms}
\end{equation}
In this equation, $z_{n_{\mathrm{cell}}}$ denotes the $z$ distance across $n_{\mathrm{cell}}$, which corresponds to the total thickness of $n_{\mathrm{cell}}$ layers, assuming the thickness of tungsten and emulsion film is constant.
$X_\mathrm{c}$ is the combined radiation length of the tungsten and emulsion film, calculated from their composition and mass fraction~\cite{Radiation:christian}.
In the \FASERnu analysis, a typical value of $X_\mathrm{c} = \SI{4.54}{\milli\metre}$ is used, but this value is changed depending on the thickness of the tungsten plates.
$s^{\mathrm{RMS}}$ is derived for each $n_{\mathrm{cell}}$.
In practice, we shift $i$ one by one and calculate $s_{i}$ for each $n_{\mathrm{cell}}$. 
If there are missing track segments, we skip the calculation of $s_{i}$ for that cell.
The measured $s^{\mathrm{RMS}}$ is a combination of the scattering $s^{\mathrm{RMS}}_{\mathrm{MCS}}\left( p, n_{\mathrm{cell}}\right)$ in \cref{eq:splanerms} and position measurement error $\sigma_{\mathrm{pos}}$. Namely,
\begin{equation}
s^{\mathrm{RMS}} = \sqrt{\left(s^{\mathrm{RMS}}_{\mathrm{MCS}}\left( p, n_{\mathrm{cell}}\right)\right)^2 + 6\,\sigma_{\mathrm{pos}}^2}.
\label{eq:srms}
\end{equation}
This equation is derived from error propagation of \cref{eq:position_difference}, assuming the interval of $z_{i}$ is constant.
We plot $s^{\mathrm{RMS}}$ as a function of $n_{\mathrm{cell}}$ for a simulated track in \cref{fig:PrecCoordFit}, where the left panel corresponds to $p_{\mathrm{true}}$ = 100 GeV and the right to 1000 GeV.
The second differences are calculated by shifting the index $i$ one step at a time, so that largely overlapping data points are reused multiple times.
As a result, the values of $s^{\mathrm{RMS}}$ are not fully statistically independent.
To account for this effect, an effective statistical weight
$w(n_{\mathrm{cell}})$ is assigned to each $s^{\mathrm{RMS}}$ value,
defined as
\begin{equation}
w(n_{\mathrm{cell}}) =
\frac{\sqrt{(n_{\mathrm{plate}}-1)/n_{\mathrm{cell}}}}
{s^{\mathrm{RMS}}}.
\label{eq:weight}
\end{equation}
This weight reflects the effective number of independent measurements
contributing to each $s^{\mathrm{RMS}}$.
The $s^{\mathrm{RMS}}$ values are then fitted with \cref{eq:srms}, and the fit result is shown as the red line.

Here, $p$ and $\sqrt{6}\,\sigma_{pos}$ are two fitting parameters.  
The momentum is derived from the dependence of the scattering term on $n_{\mathrm{cell}}^{3/2}$, which follows from the $z_{n_{\mathrm{cell}}}\sqrt{z_{n_{\mathrm{cell}}}} \propto n_{\mathrm{cell}}^{3/2}$ behavior in \cref{eq:splanerms}.
The position resolution is derived from the intercept with the $y$-axis. 
In this example, the maximum number of cell length used for fitting ($n_{\mathrm{cell}}^{\mathrm{max}}$) was set to 32.
This value, however, should be determined according to the track length and the position resolution.
Although using a larger $n_{\mathrm{cell}}$ allows to measure the effect of the scattering over a larger range and thus determining higher momenta, the effective statistical sample becomes smaller, resulting in worse resolution.
This point will be discussed in detail in the next section.

No additional correction for energy loss due to ionization and radiation of muons was applied, as MC simulation studies showed the effect to be negligible. Although muons with energies above around 100 GeV penetrating tungsten are not strictly minimum-ionizing particles, the average energy loss over 100 plates is below 0.7\% of the $p_{\mathrm{true}}$ (for example, about 3.8 GeV for a 1 TeV muon). 

\section{Evaluation of the momentum measurement performance}
\label{sec:Evaluation}
\begin{figure}[h]
    \centering
    \begin{minipage}{0.38\textwidth}
        \centering
        \includegraphics[width=\textwidth]{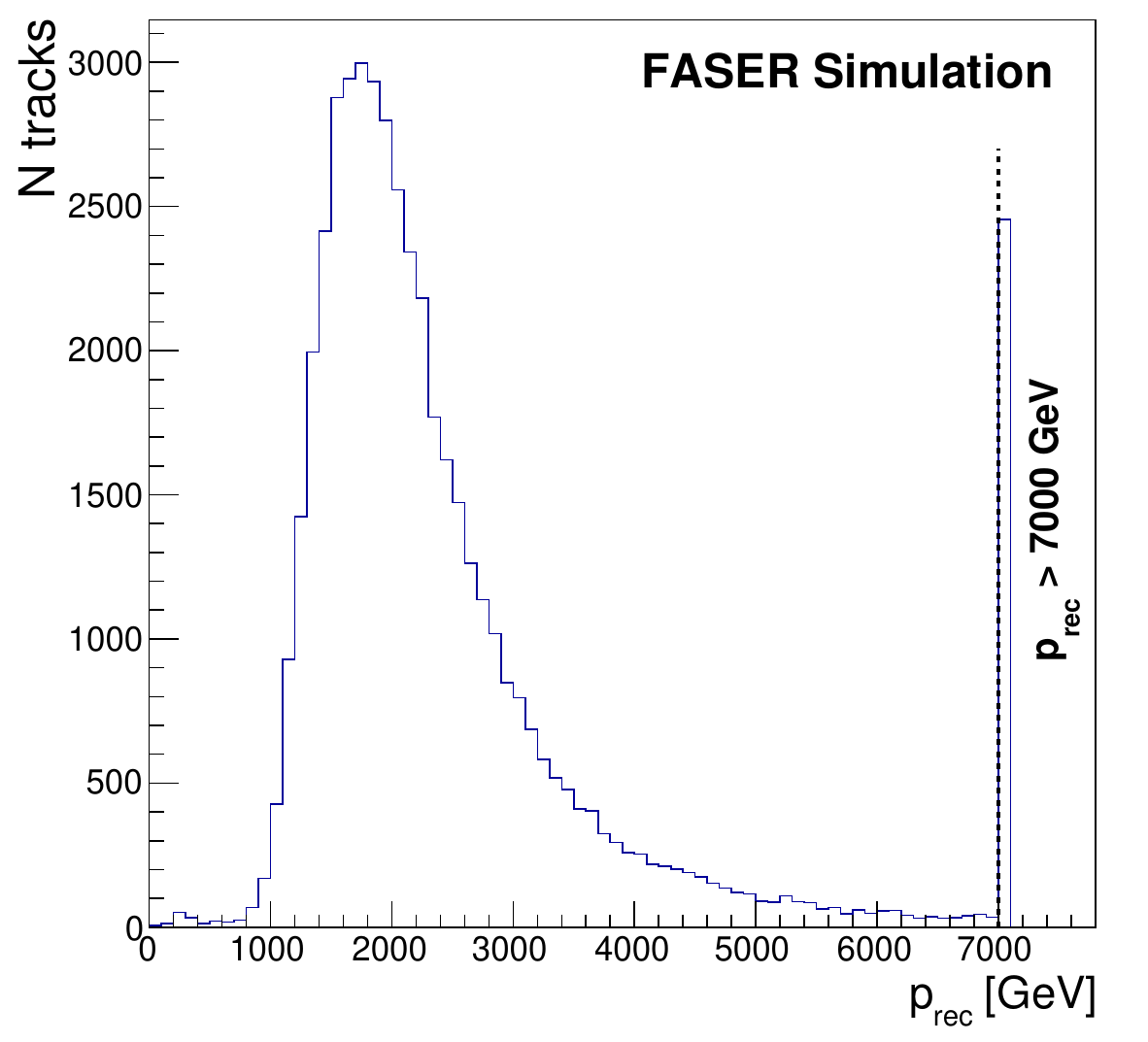}
    \end{minipage}    
    \begin{minipage}{0.38\textwidth}
        \centering
        \includegraphics[width=\textwidth]{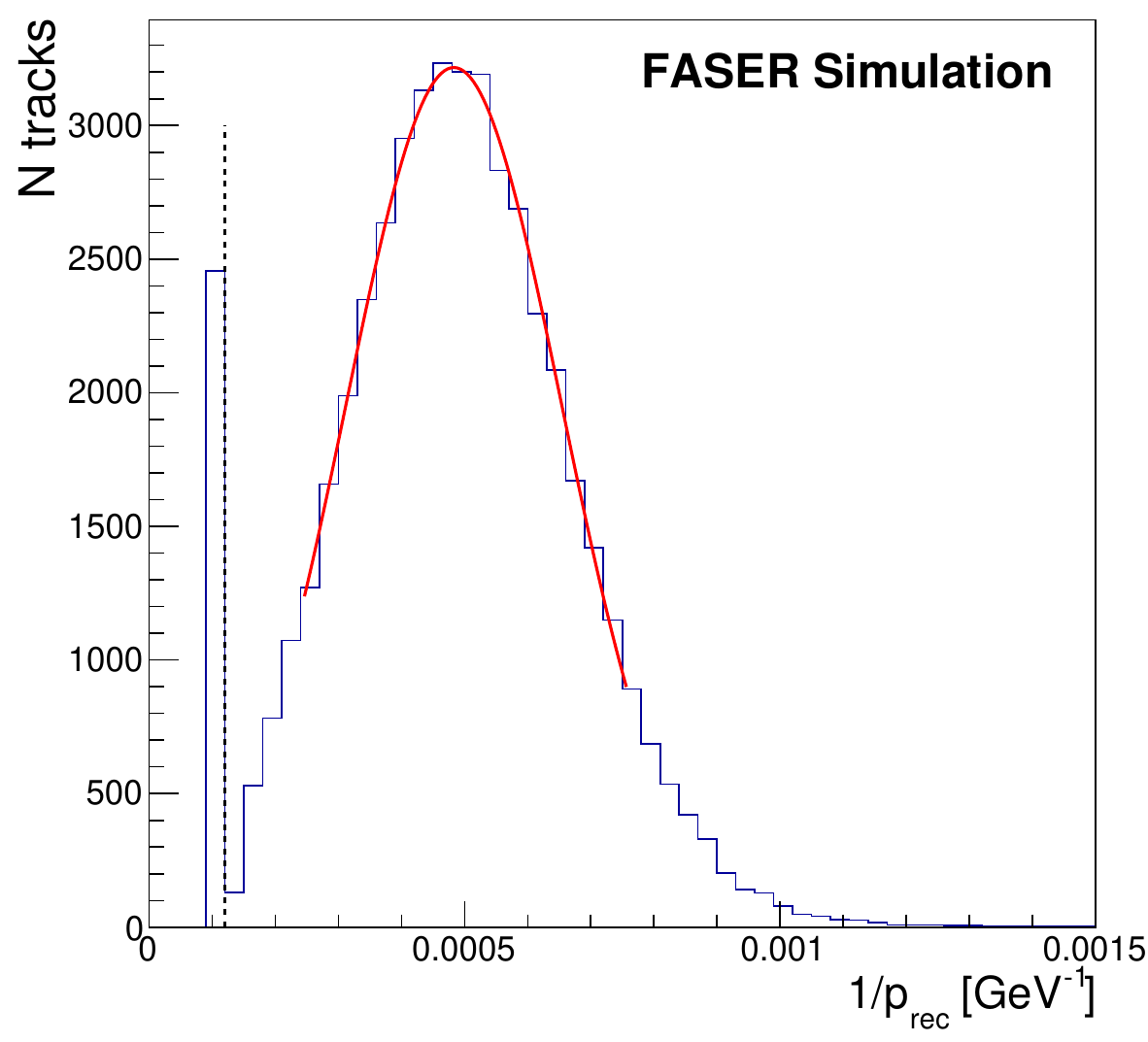}
    \end{minipage}
    \caption{Left: reconstructed momentum ($p_{\mathrm{rec}}$) distribution. Right: $1/p_{\mathrm{rec}}$ distribution. Both are for a single-energy MC sample with a true momentum of 2000 GeV. If the reconstructed momentum exceeded 7 TeV, it was set to 7 TeV. The red line in the right panel indicates a Gaussian fit performed within a range of the mean $\pm$ 1.5 RMS of the distribution, excluding tracks with $p_{\mathrm{rec}} = 7$ TeV.}
    \label{fig:MomentumEvaluationExample}
\end{figure}

The performance of the momentum measurement was evaluated using MC simulations. The muons and all other particles traversing the tungsten/emulsion detector are simulated using Geant4 \cite{Geant4:2003}, and the FTFP\_BERT physics list is used.
Single-energy samples ranging from 10 to 3000 GeV were used and a position smearing of \SI{0.3}{\micro\metre} and a base-track reconstruction efficiency of 90\% were included to account for the typical detector resolution and efficiency.
For the momentum measurement, a track length of 100 plates was used, since in the current analysis the momentum of the tracks from the neutrino-interaction candidates is measured over 100 plates \cite{FASER:2024hoe}.
Under these conditions, the optimal value of $n_{\mathrm{cell}}^{\mathrm{max}}$ is determined.
\Cref{fig:MomentumEvaluationExample} shows the reconstructed momentum ($p_{\mathrm{rec}}$) distribution (left) and the $1/p_{\mathrm{rec}}$ distribution (right) for tracks with a true momentum ($p_{\mathrm{true}}$) of 2000 GeV, using $n_{\mathrm{cell}}^{\mathrm{max}} = 24$.
If the reconstructed momentum exceeded 7 TeV, the value was set to 7 TeV. Since the LHC beam energy is about 7 TeV, it does not make sense that any particle traversing the FASER detector can have an energy higher than this.
The peak on the right side of $p_{\mathrm{rec}}$ and left side of the $1/p_{\mathrm{rec}}$ distribution corresponds to tracks whose momenta were reconstructed as 7 TeV.
The $p_{\mathrm{rec}}$ distribution shows a high-momentum tail and a mode slightly lower than $p_{\mathrm{true}}$, both resulting from the fact that the $1/p_{\mathrm{rec}}$ distribution approximately follows a Gaussian shape.
The $1/p_{\mathrm{rec}}$ distribution was fitted with a Gaussian function, where the inverse of the mean is defined as $p_{\mathrm{center}}$, and the ratio of the standard deviation to the mean ($\sigma/\mathrm{mean}$) is defined as the resolution.
In this case, $p_{\mathrm{center}}$ is 2070 GeV and the resolution is 35\%.
The fraction of tracks reconstructed with momenta below 7 TeV is defined as the success rate, which is 95\%.


\begin{figure}[tbp]
    \centering
    \includegraphics[width=0.325\textwidth, trim={0.2cm 0 0.2cm 0}, clip]{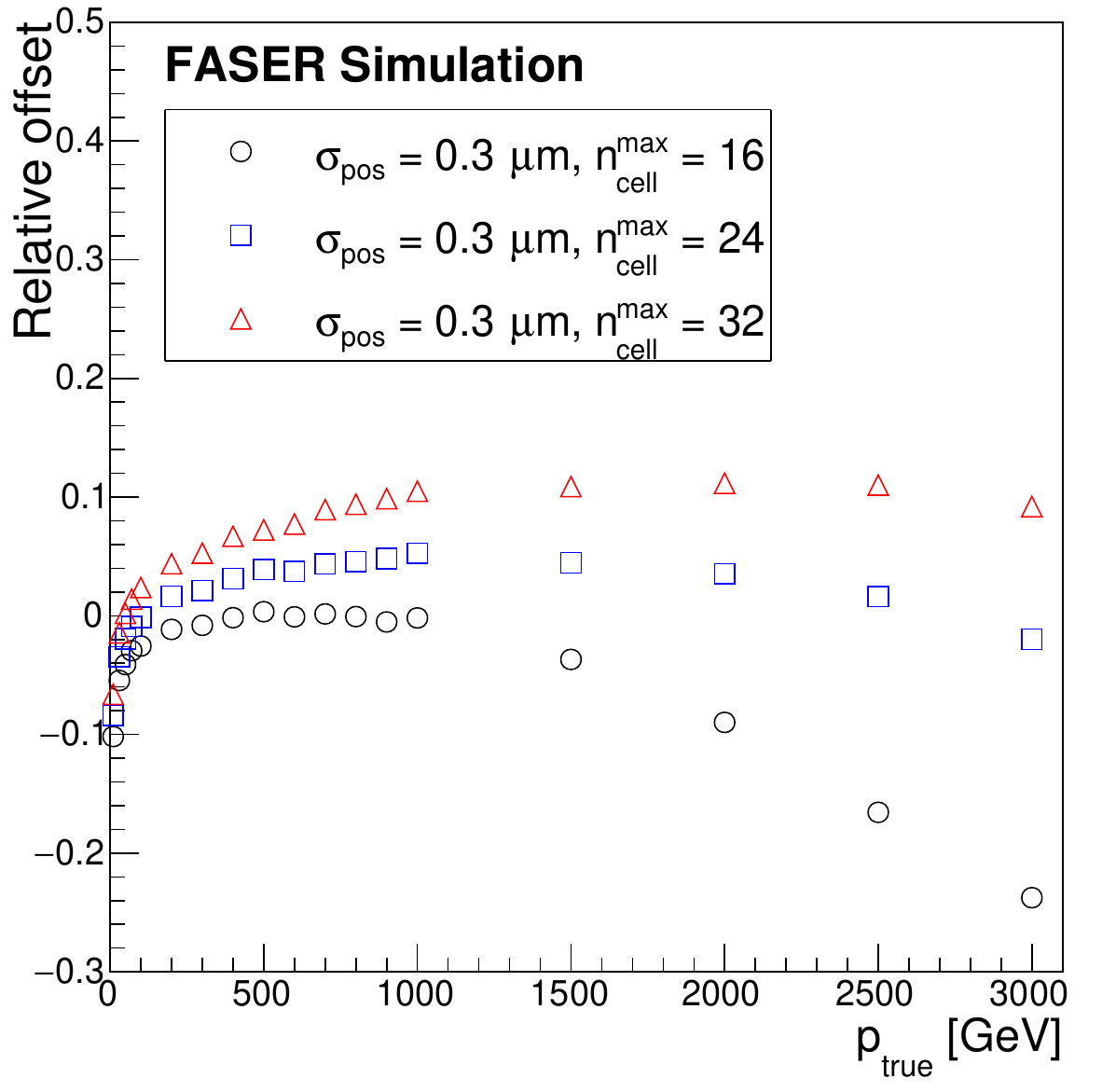}
    \includegraphics[width=0.325\textwidth, trim={0.2cm 0 0.2cm 0}, clip]{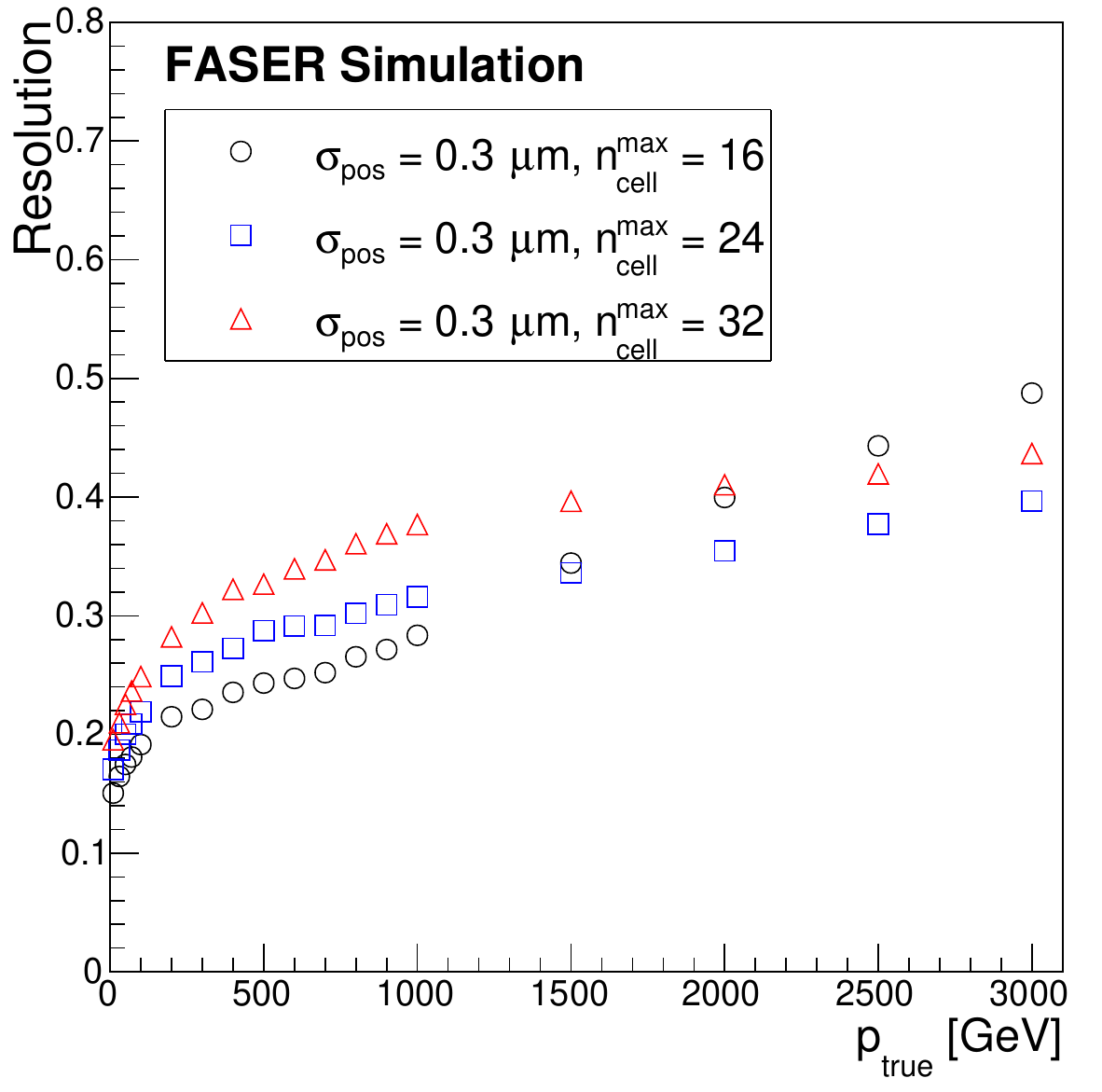}
    \includegraphics[width=0.325\textwidth, trim={0.2cm 0 0.2cm 0}, clip]{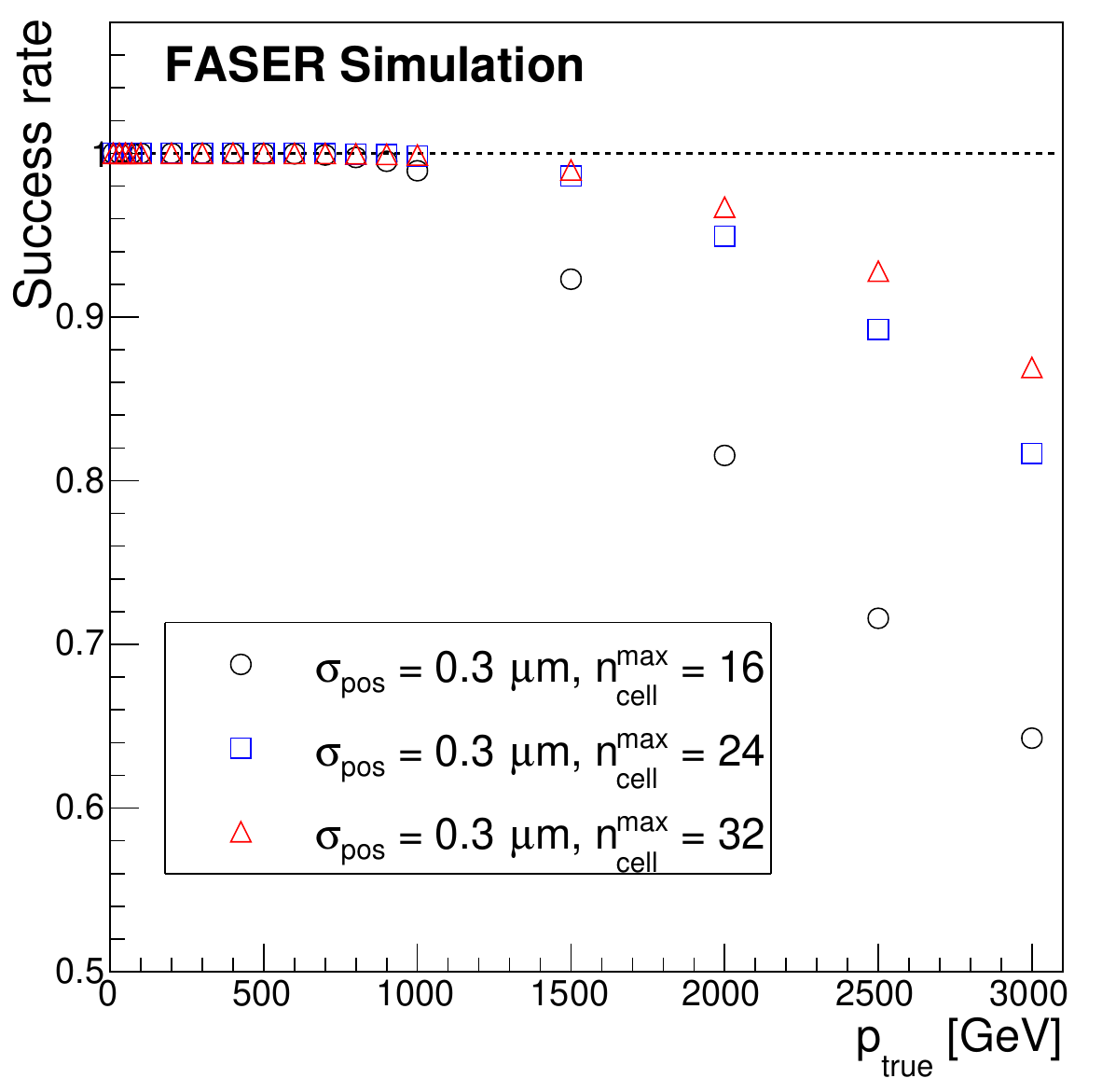}
    \caption{The relative offset (left), resolution (center), and success rate (right) as a function of $p_{\mathrm{true}}$. Black circles, blue squares, and red triangles correspond to $n_{\mathrm{cell}}^{\mathrm{max}}$ = 16, 24, and 32, respectively.
    The relative offset is defined as $(p_{\mathrm{center}} - p_{\mathrm{true}})/p_{\mathrm{true}}$, and the success rate represents the fraction of tracks reconstructed with momenta below 7 TeV for each $p_{\mathrm{true}}$.}
    \label{fig:EvaluationDifferentNcellMax}
\end{figure}

\begin{figure}[h]
    \centering
    \includegraphics[width=0.5\linewidth]{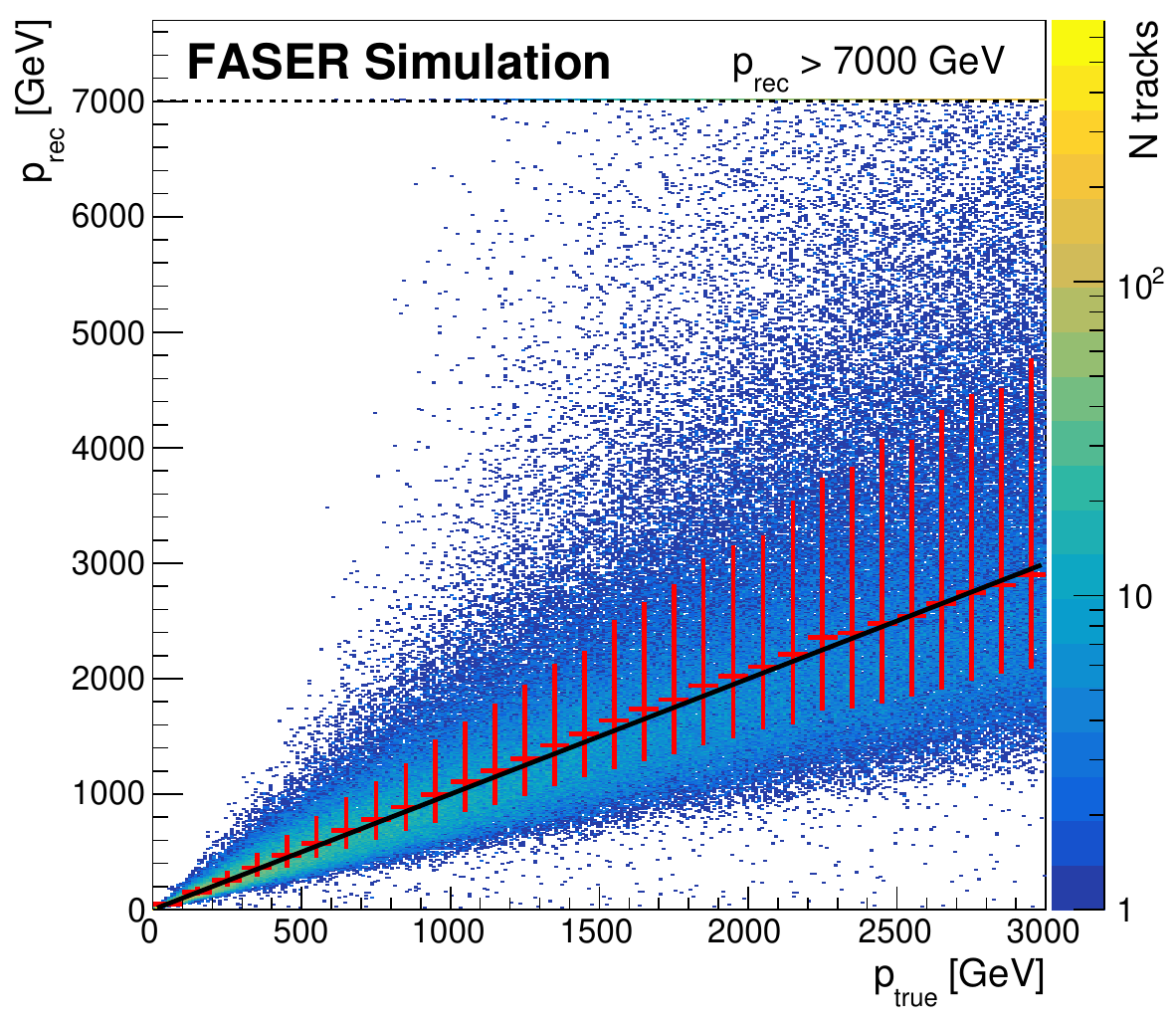}
    \caption{$p_{\mathrm{rec}}$ as a function of $p_{\mathrm{true}}$ using $n_{\mathrm{cell}}^{\mathrm{max}}$ = 24. The MC simulation assumes a flat momentum distribution from 1 GeV to 3000 GeV. The colour scale represents the number of entries in each bin. The black line indicates $p_{\mathrm{rec}} = p_{\mathrm{true}}$, and the red profile histograms show the $p_{\mathrm{center}}$ values with 1$\sigma$ uncertainties for each 100 GeV bin.}
    \label{fig:PerformanceNcell24}
\end{figure}

\begin{figure}[h]
    \centering
    \includegraphics[width=0.325\textwidth, trim={0.2cm 0 0.2cm 0}, clip]{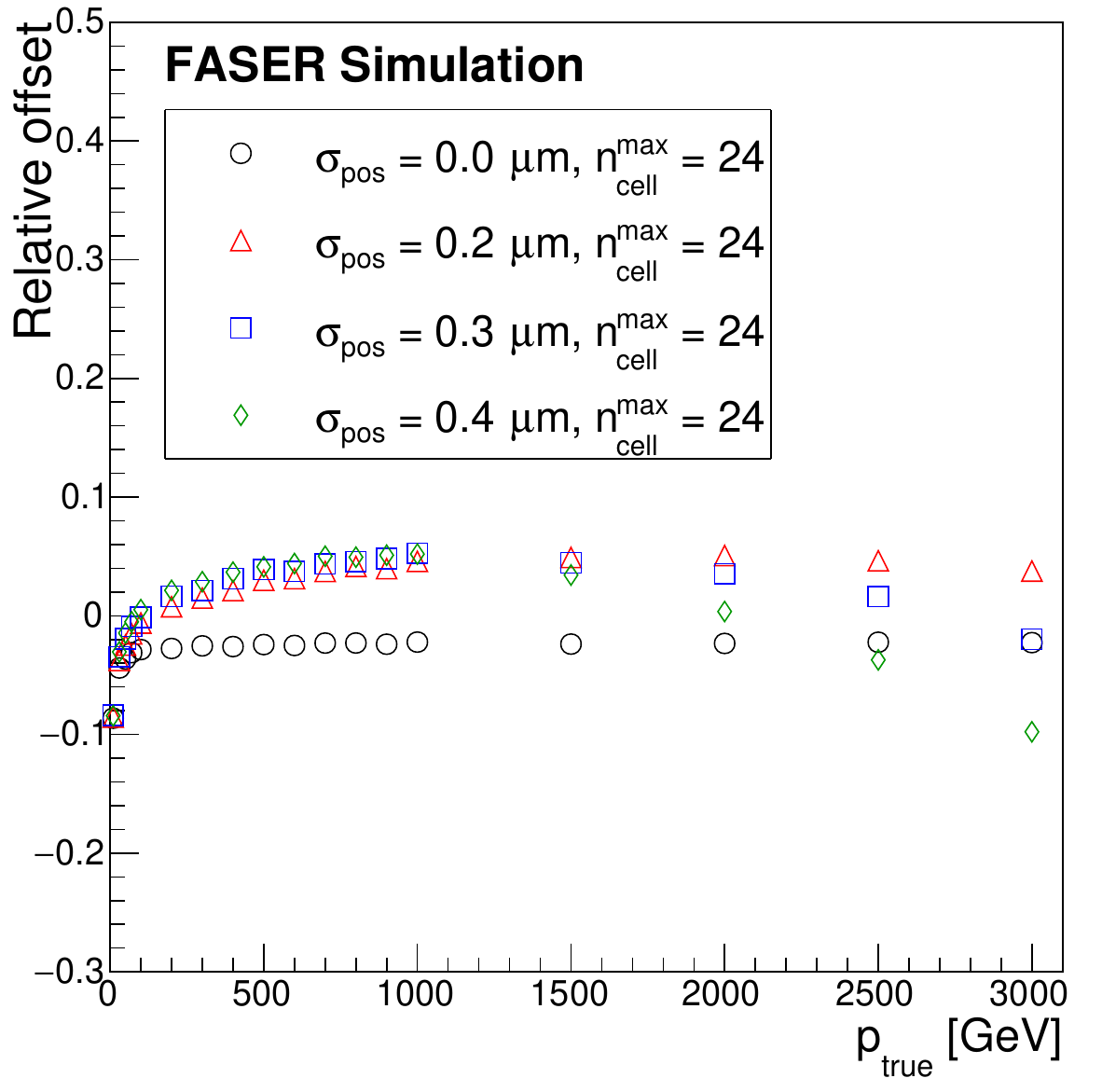}
    \includegraphics[width=0.325\textwidth, trim={0.2cm 0 0.2cm 0}, clip]{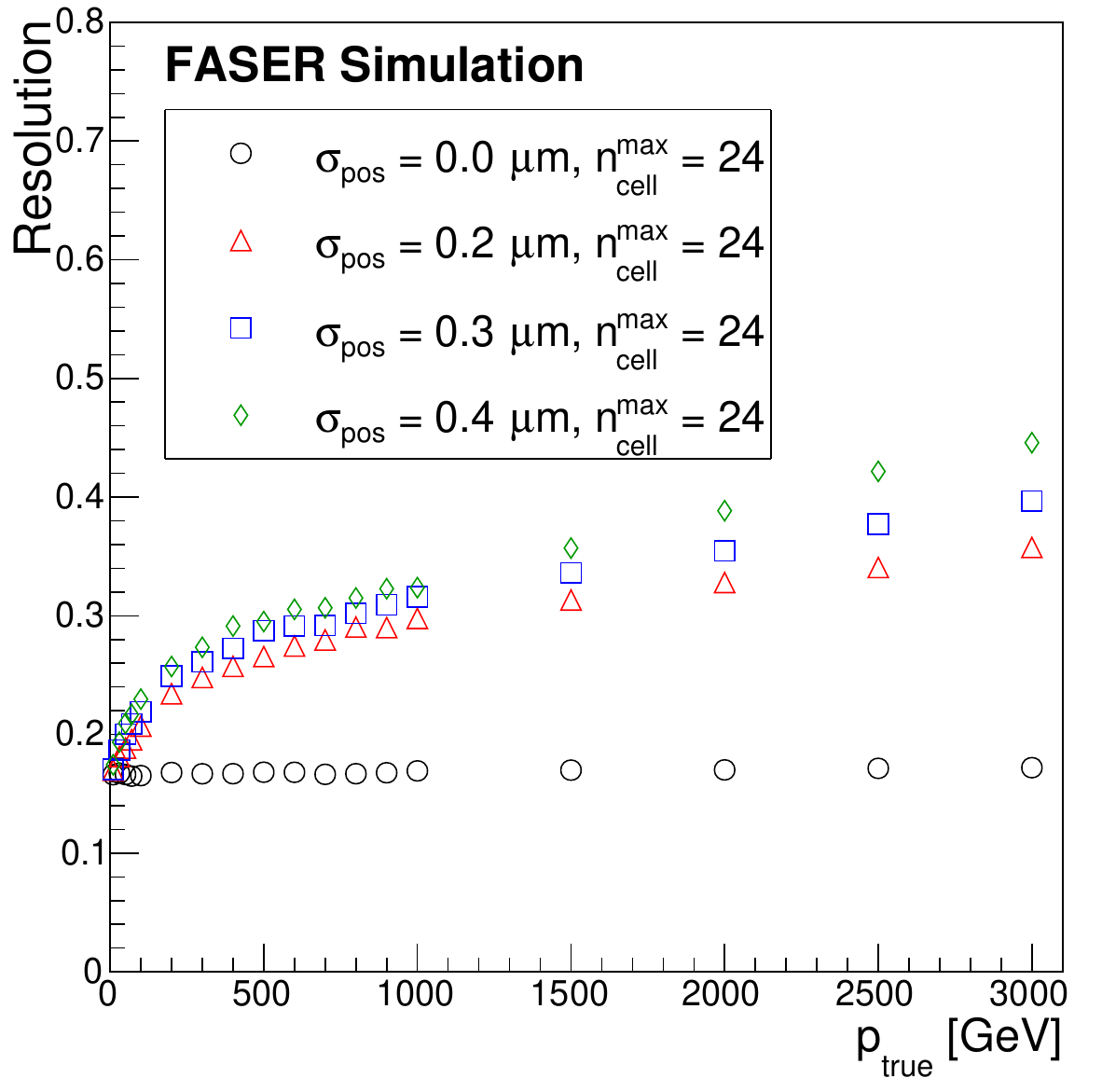}
    \includegraphics[width=0.325\textwidth, trim={0.2cm 0 0.2cm 0}, clip]{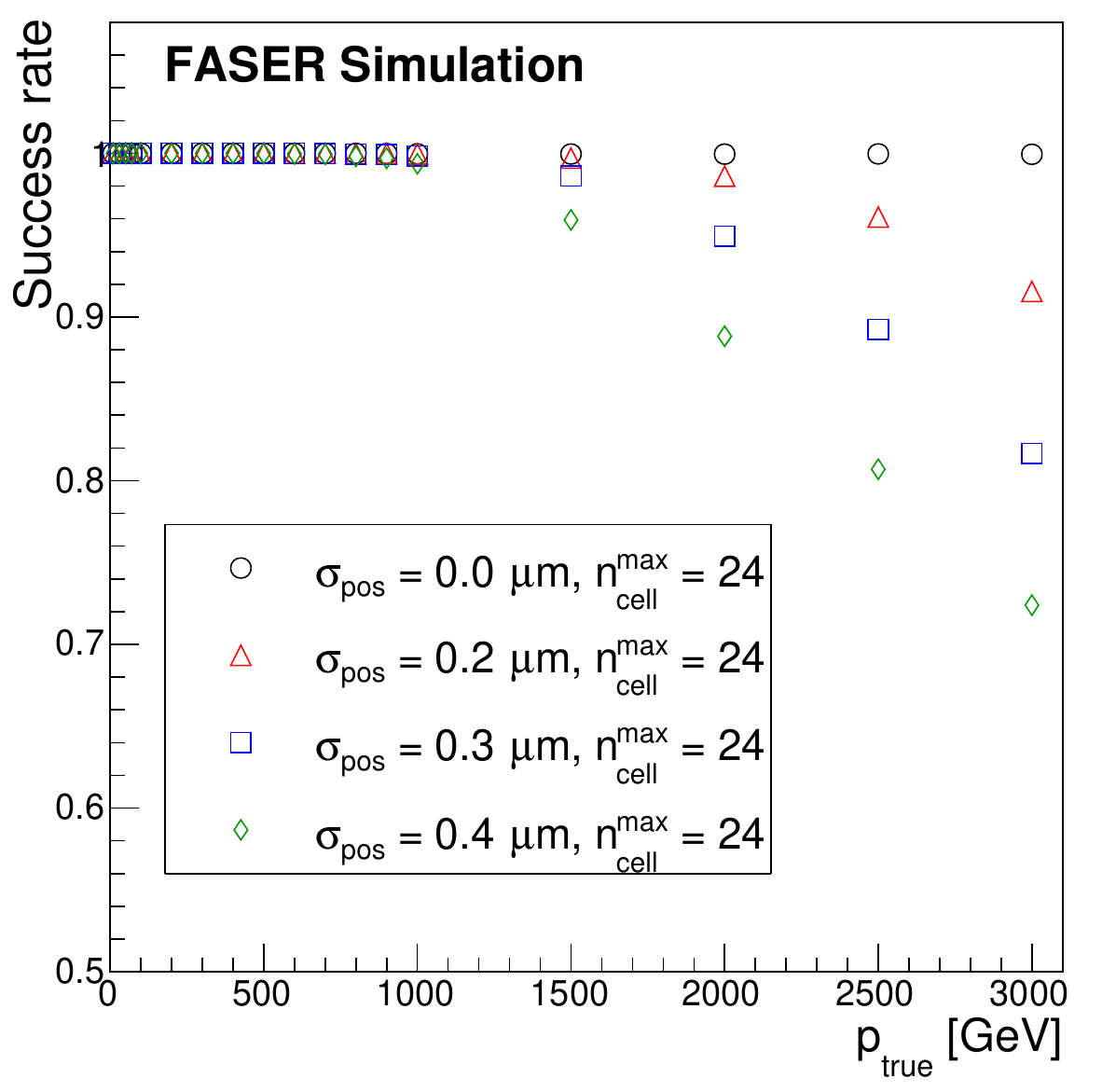}
    \caption{The relative offset (left), resolution (center), and success rate (right) as a function of $p_{\mathrm{true}}$. Black circles, red triangles, blue squares and green diamonds correspond to $\sigma_{\mathrm{pos}}$ = 0.0, 0.2, 0.3 and 0.4 \SI{}{\micro\metre}. $n_{\mathrm{cell}}^{\mathrm{max}}$ is fixed at 24.
    The relative offset is defined as $(p_{\mathrm{center}} - p_{\mathrm{true}})/p_{\mathrm{true}}$, and the success rate represents the fraction of tracks reconstructed with momenta below 7 TeV for each $p_{\mathrm{true}}$.}
    \label{fig:EvaluationDifferentSigmaPos}
\end{figure}

\Cref{fig:EvaluationDifferentNcellMax} shows the relative offset (left), resolution (center), and success rate (right) as a function of $p_{\mathrm{true}}$.
The relative offset is defined as $(p_{\mathrm{center}} - p_{\mathrm{true}})/p_{\mathrm{true}}$.
Black circles, blue squares, and red triangles correspond to $n_{\mathrm{cell}}^{\mathrm{max}}$ = 16, 24, and 32, respectively. The values of $n_{\mathrm{cell}}^{\mathrm{max}}$ are chosen to correspond to typical track lengths of approximately one-sixth, one-quarter, and one-third of the 100 plates used for the momentum measurement, in order to ensure sufficient statistical independence. 
For $n_{\mathrm{cell}}^{\mathrm{max}}$ = 16, the resolution is the best among the three conditions below 1 TeV; The relative offset becomes large above 1 TeV, and the success rate decreases significantly.
Conversely, $n_{\mathrm{cell}}^{\mathrm{max}}$ = 32 provides the highest success rate, but the resolution is the worst.
This is because using a larger $n_{\mathrm{cell}}$ allows the effects of scattering to be measured over a longer distance, improving sensitivity to the small scattering of higher-momentum particles.
However, the total number of plates is limited to 100, which reduces the statistical sample and results in a broader $s^{\mathrm{RMS}}$ distribution.
The case of $n_{\mathrm{cell}}^{\mathrm{max}}$ = 24 shows a small relative offset, a resolution intermediate between those of 16 and 32, and a success rate comparable to that of 32.
Based on these results, $n_{\mathrm{cell}}^{\mathrm{max}}$ = 24 is adopted as the baseline for the momentum measurement.
\Cref{fig:PerformanceNcell24} shows the performance of the momentum measurement using $n_{\mathrm{cell}}^{\mathrm{max}}$ = 24.
The MC simulation assumes a flat momentum distribution from 1 GeV to 3000 GeV.
The red profile histograms show the $p_{\mathrm{center}}$ values with their 1$\sigma$ Gaussian width for each 100 GeV bin. The widths are obtained by propagating the sigma of the Gaussian fit of $1/p_{\mathrm{rec}}$ into momentum space.
The black line corresponds to $p_{\mathrm{rec}} = p_{\mathrm{true}}$.
As seen in the figure, the developed method maintains good linearity even above 1 TeV. 
\Cref{fig:EvaluationDifferentSigmaPos} shows the relative offset (left), resolution (center), and success rate (right) as functions of $p_{\mathrm{true}}$. The value of $n_{\mathrm{cell}}^{\mathrm{max}}$ is fixed at 24, while several $\sigma_{\mathrm{pos}}$ (0.0, 0.2, 0.3 and 0.4 \SI{}{\micro\metre}) are compared. For $\sigma_{\mathrm{pos}}$ = \SI{0.0}{\micro\metre}, the relative offset, resolution and success rate remain almost constant. This demonstrate that there is no momentum dependent bias when perfect alignment is assumed, and the theoretical resolution is around 17\%. As $\sigma_{\mathrm{pos}}$ increases, all three quantities degrade. However, the change in the relative offset can be explained by the distortion of the $s^{\mathrm{RMS}}$ distribution. The deviation of the relative offset among the three $\sigma_{\mathrm{pos}}$ conditions is sufficiently smaller than their resolutions, and thus this effect can be negligible.
In this MC study, the beam is incident perpendicular to the plates; in addition, tracks with different angles with respect to the plates were also checked.
Muon tracks with energies above 100 GeV from neutrino interactions are largely contained within $\tan\theta < 0.1$.
Although the position resolution degrades with increasing track angle, within this angular range, it remains at approximately \SI{0.35}{\micro\metre}, and therefore has no practical impact on $p_{\mathrm{center}}$.
In addition, the momentum measurement was performed independently using only the $x$ or only the $y$ information, and consistent $p_{\mathrm{center}}$ values and resolutions were obtained.



\section{Analysis of muon test beam data}
\label{sec:Testbeam}

In this section, we compare the momentum measurement results obtained from the muon test beam data with those from MC simulations.


\subsection{Test beam experiment}
\begin{figure}[h]
    \centering
    \begin{minipage}{0.46\textwidth}
        \centering        
        \includegraphics[width=\textwidth]{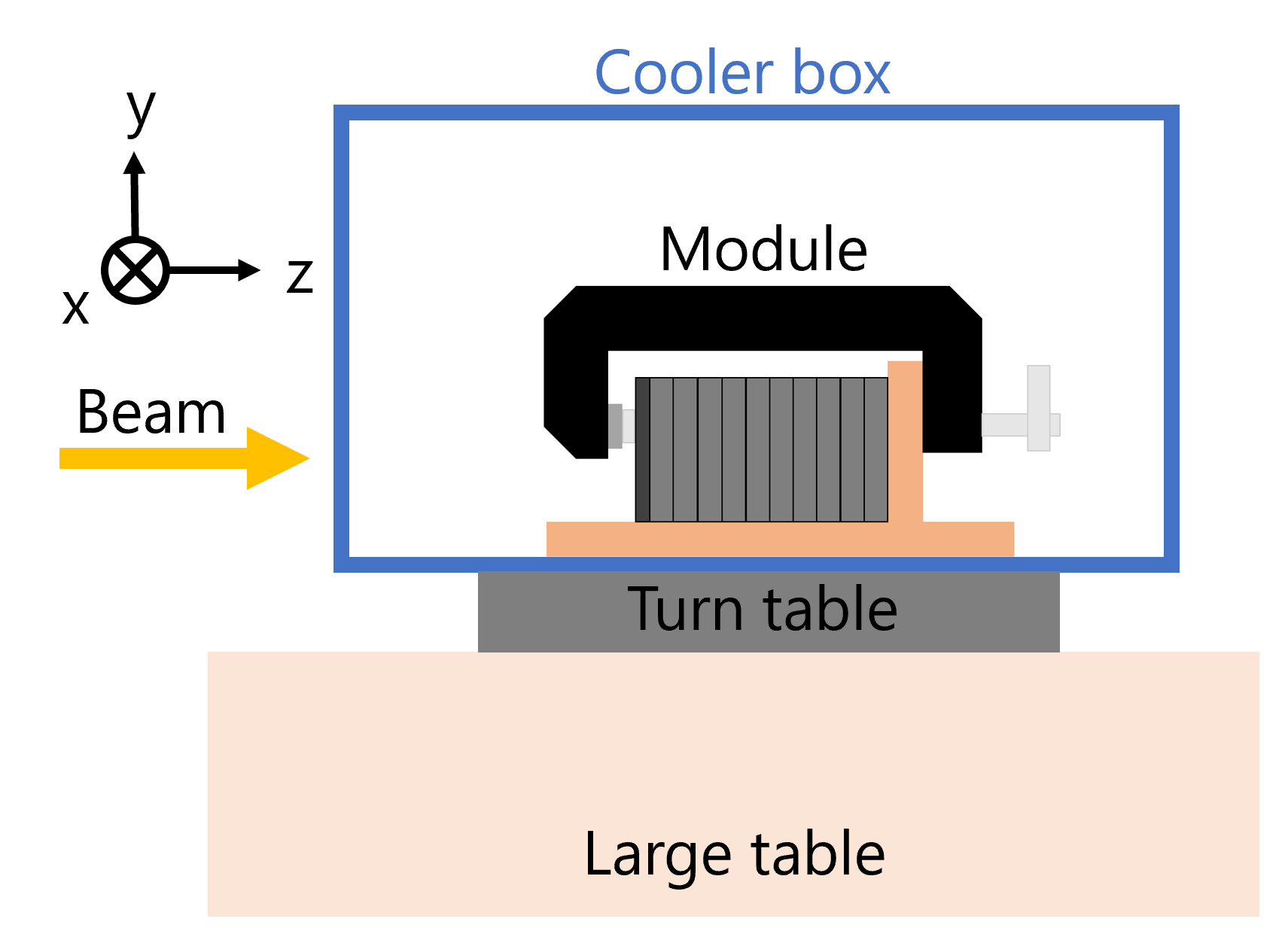}
    \end{minipage}
    \begin{minipage}{0.48\textwidth}
        \centering
        \includegraphics[width=\textwidth]{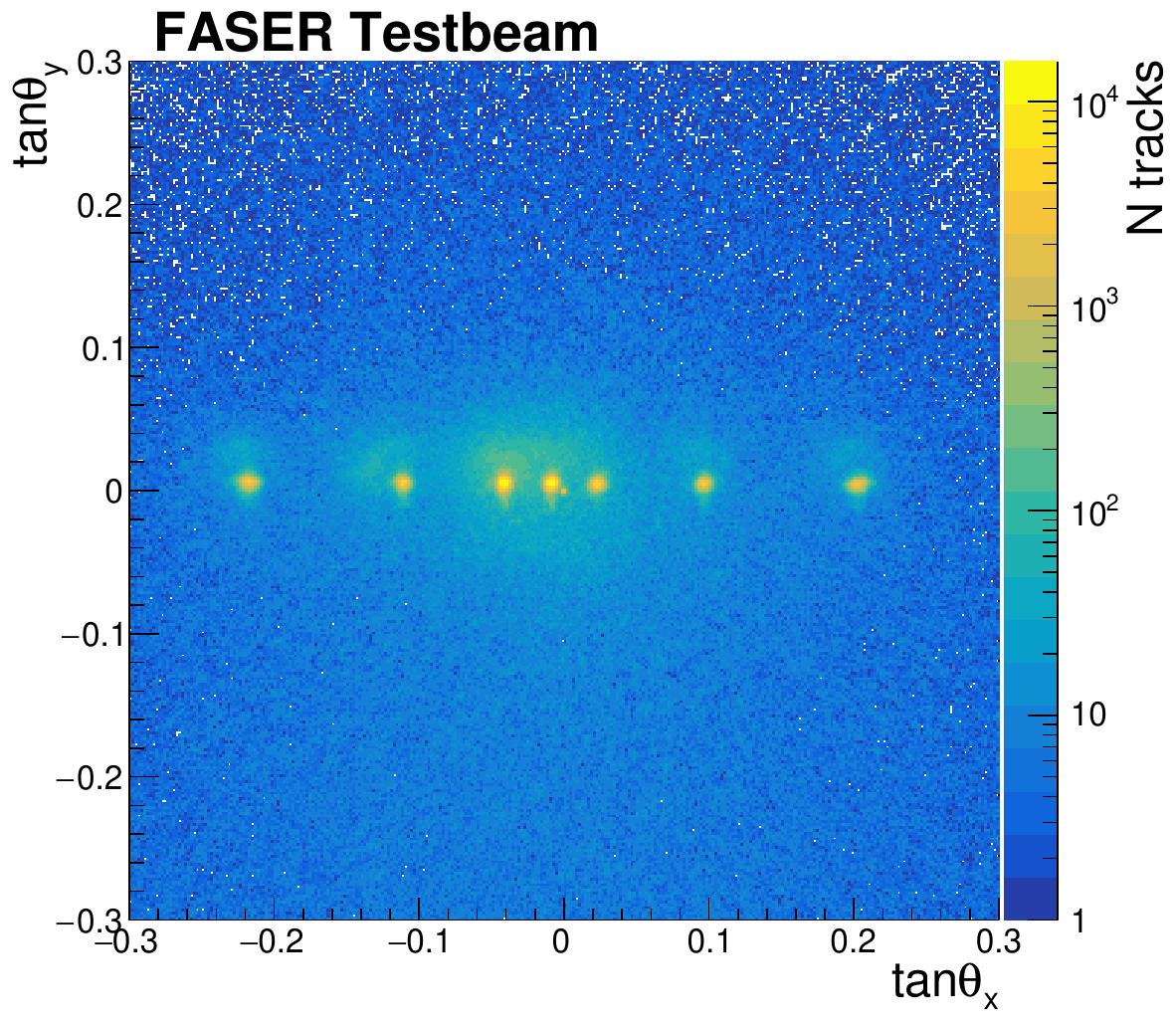}
    \end{minipage}
    \caption{Left: Setup of the irradiation to the module. The z axis is approximately aligned with the beam direction, and the x–y plane is perpendicular to the beam. Right: Base-track two dimensional angle distribution of the most upstream film.}
    \label{fig:BaseTrack2DAngleDistribution}
\end{figure}

The test beam was performed at the H8 beamline of CERN’s SPS between July 3rd and 10th, 2024, to collect the data for validating the momentum measurement method. 
The irradiation was performed on July 6th, and the process took approximately five hours from the start of the first beam exposure to the end of the last. 
The emulsion detector module consisted of 100 plates of emulsion films alternating with 1.1 mm-thick tungsten plates. 
The module had dimensions of 12.5 cm x 10.0 cm x 14.5 cm (x, y, z) and a total mass of 25 kg.
As shown in \cref{fig:BaseTrack2DAngleDistribution} (left), 
the coordinate system is defined such that the z-axis is approximately aligned with the beam direction, while the x–y plane is perpendicular to the beam. 
The detector was composed of ten sub-modules, each made by vacuum-packing ten interleaved emulsion films and tungsten plates in an air tight envelope, as is done for the \FASERnu detector.
All sub-modules were fixed together with a clamp and placed inside a cooler box, which was set on top of a turntable.
The detector was irradiated with muon beams of 100, 200, and 300 GeV at various incident angles by rotating the turntable. 
The configuration for the different irradiations is summarized in \cref{tab:IrradiationT201}. 

\begin{table}[h]
    \centering
    \small
    \begin{tabular}{|c|c|c|c|}
    \hline
        particle & Momentum (GeV) & tan$\theta_x$ & Expected density ($\mu^{-}$/cm$^2$)  \\
    \hline
    \hline
        $\mu^-$ & 100 & +0.03 & 3.0 x 10$^3$ \\
    \hline
        $\mu^-$ & 200 & -0.03 & 3.0 x 10$^3$ \\
    \hline
        $\mu^-$ & 200 & -0.1, -0.2 & 1.0 x 10$^3$ \\
    \hline
        $\mu^-$ & 300 & 0.0 & 3.0 x 10$^3$ \\
    \hline
        $\mu^-$ & 300 & +0.1, +0.2 & 1.0 x 10$^3$ \\
    \hline
    \end{tabular}
    \caption{Summary of the different muon beam exposure configurations}
    \label{tab:IrradiationT201}
\end{table}

\subsection{Data analysis}
\begin{figure}[tbp]
    \centering
    \begin{minipage}{0.32\textwidth}
        \centering
        \includegraphics[width=\textwidth]{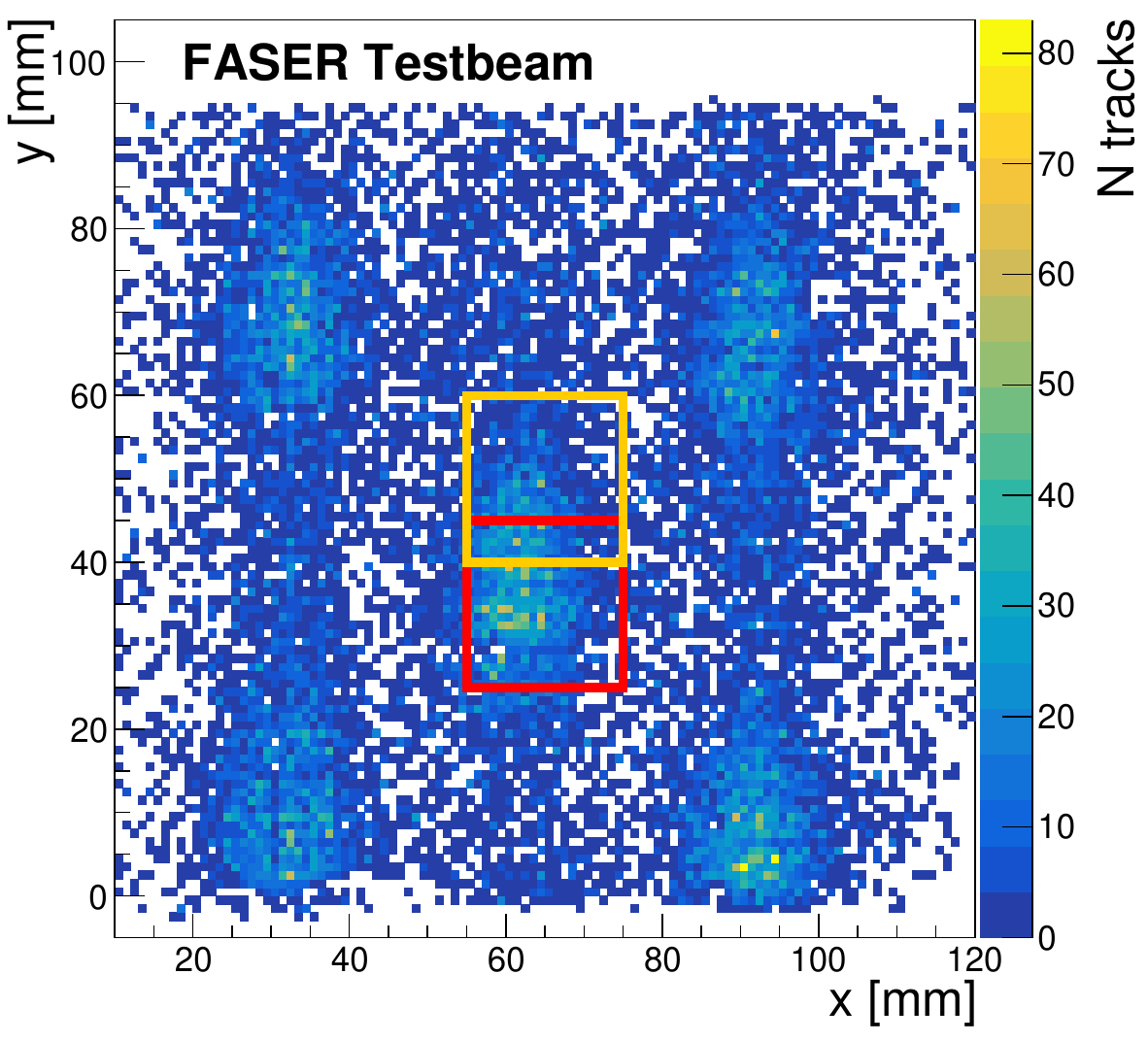}
    \end{minipage}
    \begin{minipage}{0.32\textwidth}
        \centering
        \includegraphics[width=\textwidth]{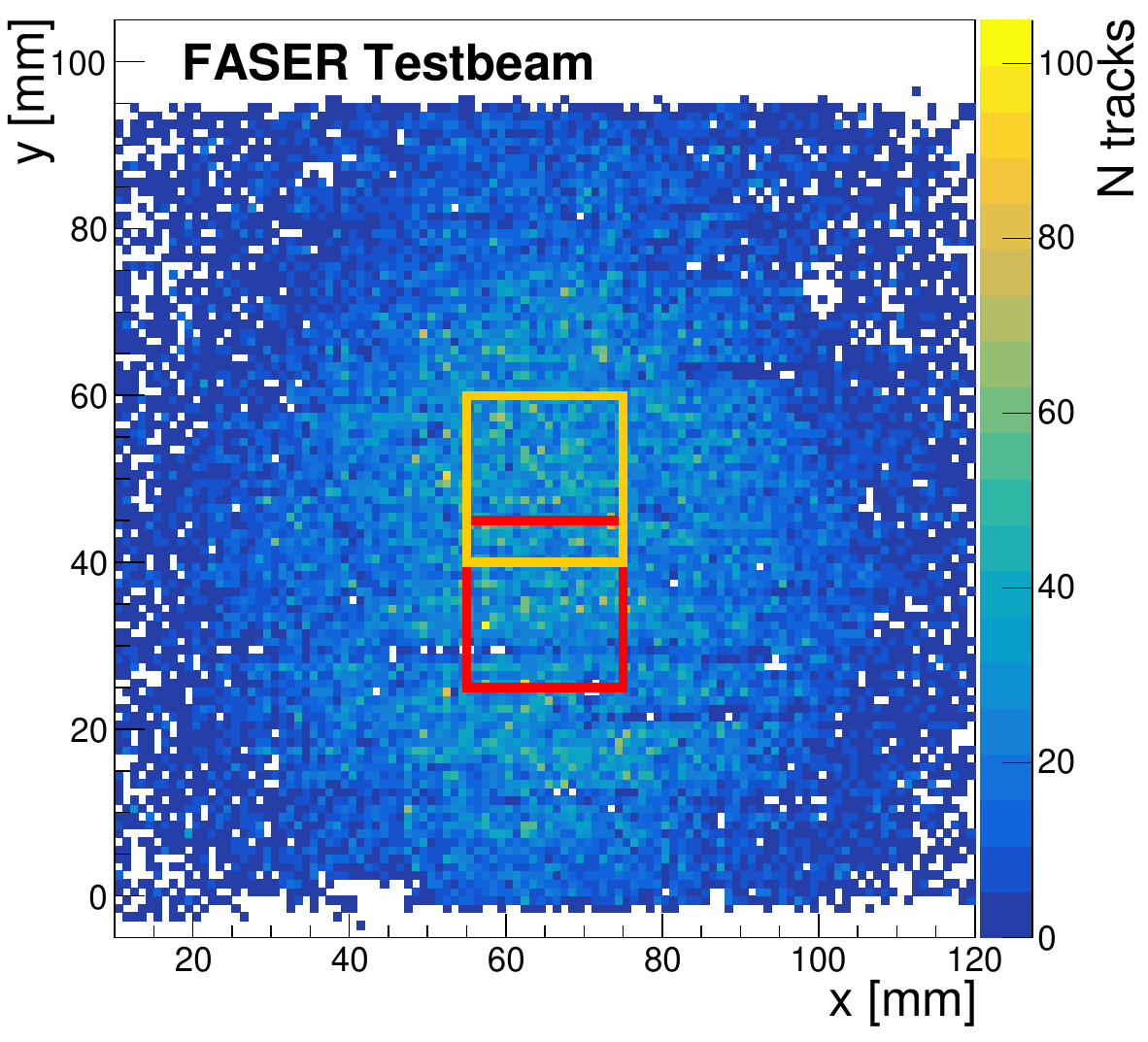}
    \end{minipage}
    \begin{minipage}{0.32\textwidth}
        \centering
        \includegraphics[width=\textwidth]{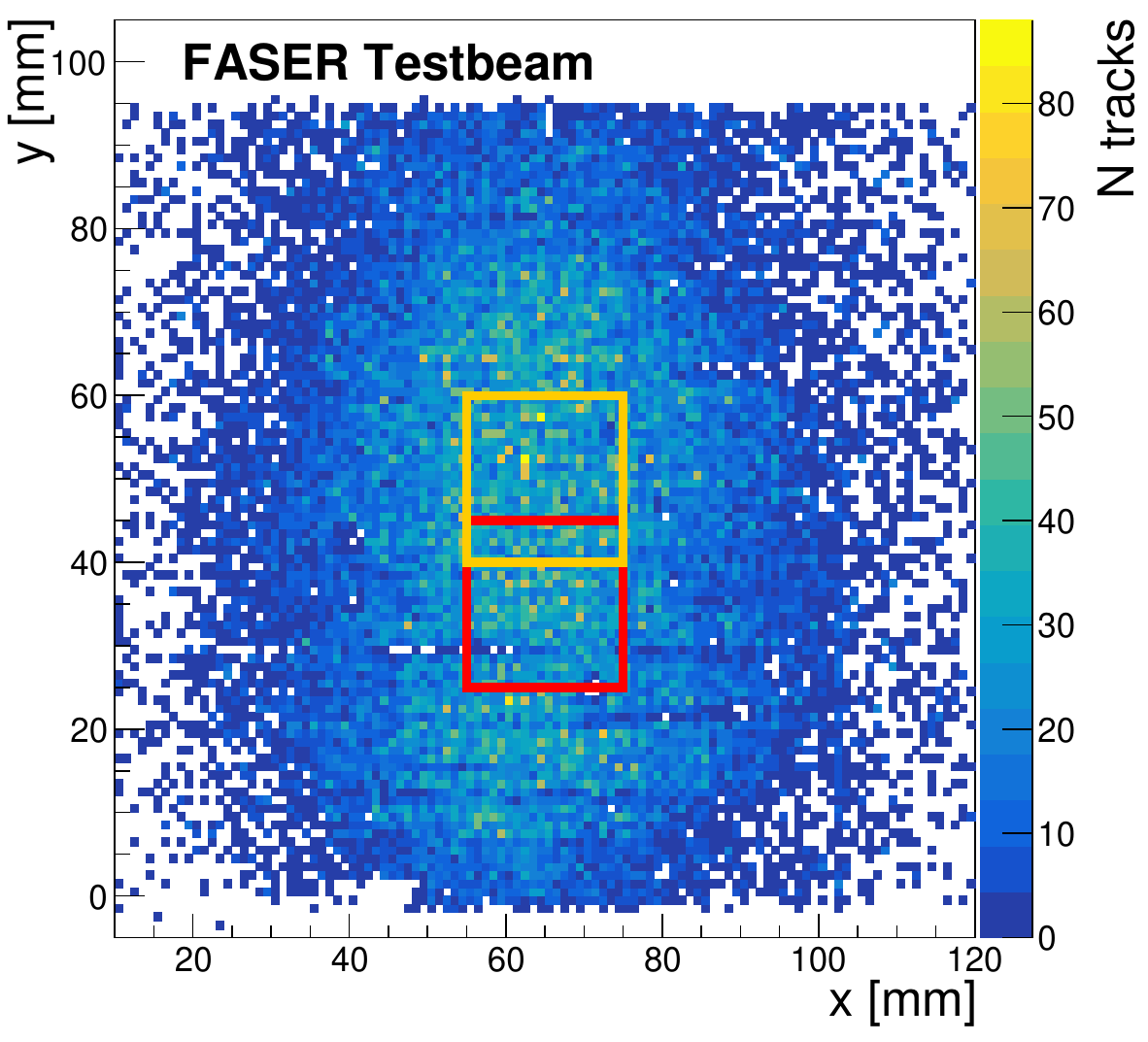}
    \end{minipage}
    \caption{
    Base-track hit position distribution of the most upstream film around the three central angular peaks shown in \cref{fig:BaseTrack2DAngleDistribution}. Left: 100 GeV. Center: 200 GeV. Right: 300 GeV. The 100 GeV beam was irradiated at five positions. The red and orange squares indicate area A and area B, respectively, centered at [x, y] = [65, 35] mm and [x, y] = [65, 50] mm. The two areas overlap by 5 mm in the y direction.}
    \label{fig:BaseaTrackHitDistribution}
\end{figure}

The tracks recorded in the irradiated films were read out using the HTS \cite{Yoshimoto:2017ufm} system. The collected data were reconstructed using the same reconstruction procedure for the \FASERnu detector \cite{FASERnu:Reconstruction}. 
\Cref{fig:BaseTrack2DAngleDistribution} (right) shows the two-dimensional base-track angular distribution in the x-y plane of the most upstream film. Seven angular peaks in the x-direction were clearly observed. \Cref{fig:BaseaTrackHitDistribution} shows the hit distributions for the three angular peaks near tan$\theta_x$ = 0, corresponding to 100 GeV (left), 200 GeV (center) and 300 GeV (right). The 100 GeV beam had a narrower profile compared to the 200 GeV and 300 GeV beams, and it was irradiated at five different spots. In \cref{fig:BaseaTrackHitDistribution}, the red and orange boxes indicate the two analyzed sub-areas of 2 cm by 2 cm, defined as area A and area B, respectively. Area A is centered at [x, y] = [65, 35] mm, and area B at [65, 50] mm, with an overlap of 5 mm in the y direction. These regions were chosen because they contain all three beam momenta. 

\begin{figure}[!b]
    \centering
    \begin{minipage}{0.48\textwidth}
        \centering
        \includegraphics[width=\textwidth]{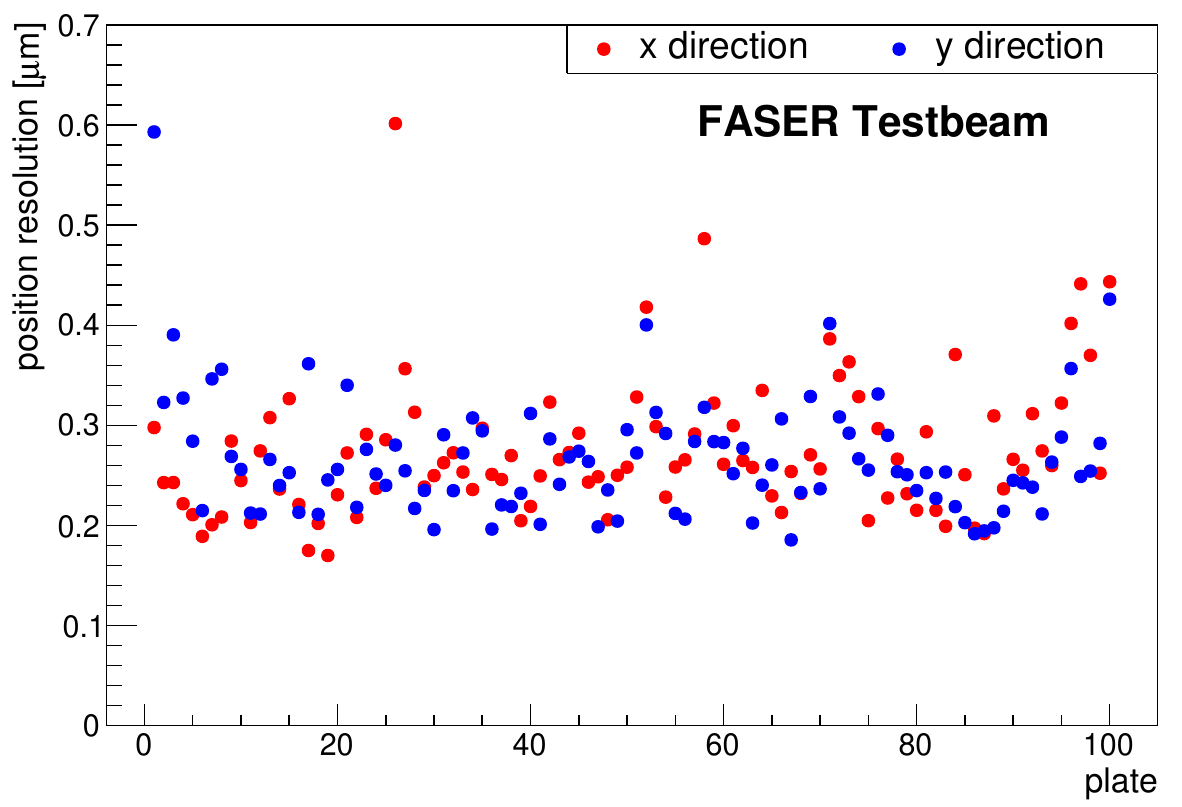}
    \end{minipage}
    \begin{minipage}{0.48\textwidth}
        \centering
        \includegraphics[width=\textwidth]{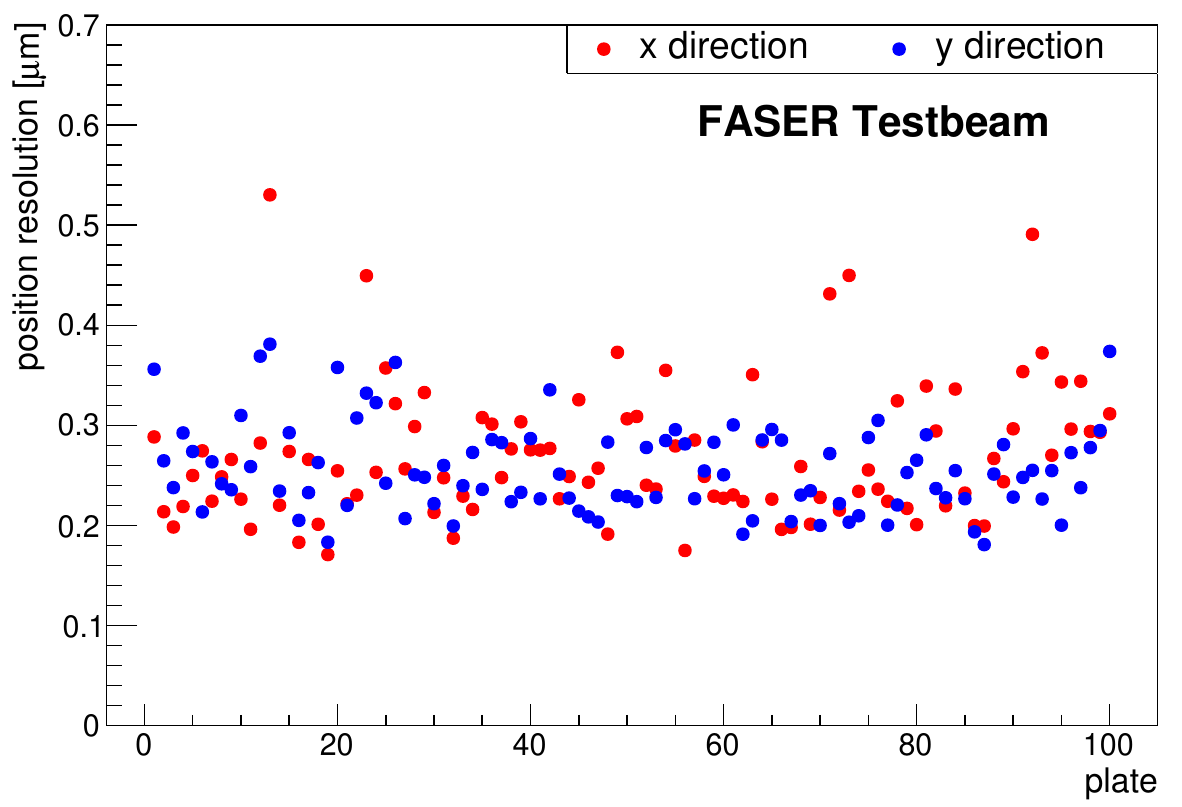}
    \end{minipage}
    \caption{Position resolution for each film in the two analysis areas. Area A (left) and area B (right) are centered at [x, y] = [65, 35] mm and [65, 50] mm, respectively. Red and blue dots show the x and y directions, respectively. The same calculation method as in \cite{FASERnu:Reconstruction} was used.}
    \label{fig:PositionResolution}
\end{figure}

For each of the two sub-areas, 100 plates were reconstructed, and the data quality was subsequently evaluated. Here, we discuss the single-film efficiency and position resolution, and the definitions of the calculation methods are given in \cite{FASERnu:Reconstruction}. 
The average single-film efficiency was measured to be 93.6\% in area A and 94.2\% in area B, respectively. 
\Cref{fig:PositionResolution} presents the position resolution in the x and y directions for each plate in the two areas. 
The left and right panels correspond to area A and area B, respectively, and the average value was \SI{0.27}{\micro\metre}. The poorer position resolution observed in certain plates is likely due to local distortions of the emulsion films. 
Since this method evaluates the displacement at the center of five consecutive plates, it overestimates the effective single-plate position resolution.
The appropriate single-plate resolution is to be obtained by multiplying \SI{0.27}{\micro\metre} by a factor of $\sqrt{4/5}$; therefore, a position smearing of \SI{0.24}{\micro\metre} is used in the MC simulation.
The data quality in terms of efficiency and resolution is comparable to that of the \FASERnu detector, indicating that they are suitable for validating the momentum measurement method for \FASERnu.


\begin{figure}[b]
    \centering
    \includegraphics[width=0.96\linewidth]{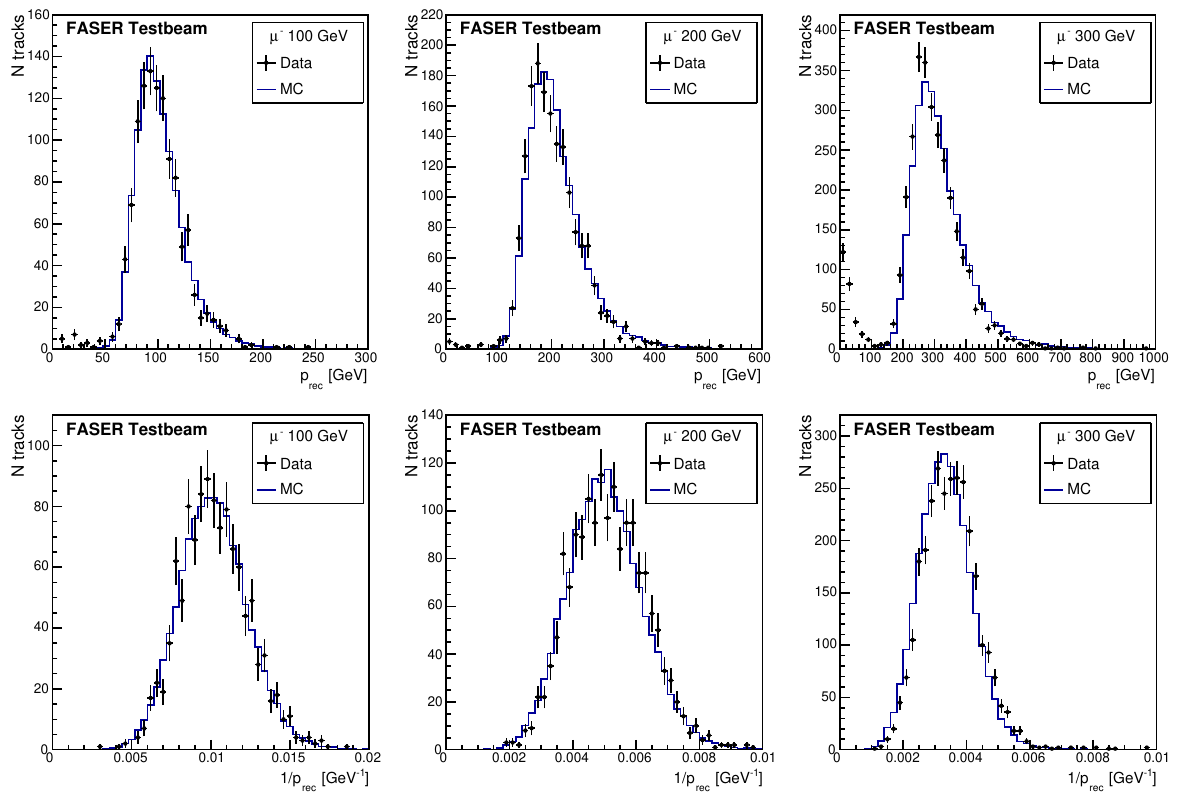}
    \caption{Comparison of $p_{\mathrm{rec}}$ (top) and $1/p_{\mathrm{rec}}$ (bottom) distributions between test beam data and MC simulation. Black points and blue histogram represent the data and the MC, respectively. From left to right: 100, 200, and 300 GeV samples are shown. The MC distributions are normalized to the total number of data tracks for 100 and 200 GeV, and to data tracks with reconstructed momenta above 100 GeV for 300 GeV.}
    \label{fig:MomentumMeasurementResult}
\end{figure}

\begin{table}[!b]
    \centering
    \small
    \begin{tabular}{|c|c|c|c|c|}
    \hline
        Beam momentum & $p_{\mathrm{center}}$ [GeV] (Data) & $p_{\mathrm{center}}$ [GeV] (MC) & Resolution [\%] (Data) & Resolution [\%] (MC) \\ \hline 
        100 GeV & 98.1$^{+4.6}_{-4.3}$ & 98.4$^{+0.1}_{-0.1}$ & 20.7$\pm$0.6 & 20.7$\pm$0.1 \\ \hline
        200 GeV & 195.2$^{+9.3}_{-8.5}$ & 198.7$^{+0.2}_{-0.2}$ & 22.7$\pm$0.6 & 22.6$\pm$0.1 \\ \hline
        300 GeV & 286.6$^{+13.5}_{-12.4}$ & 299.7$^{+0.4}_{-0.3}$ & 23.2$\pm$0.4 & 24.2$\pm$0.1 \\ \hline
    \end{tabular}
    \caption{Summary of $p_{\mathrm{center}}$ values and resolution for the data and MC. From top to bottom: 100, 200, and 300 GeV.}
    \label{tab:ComparePcenterDataToMC}
\end{table}

\begin{figure}[b]
    \centering
    \begin{minipage}{0.48\textwidth}
        \centering
        \includegraphics[width=\textwidth]{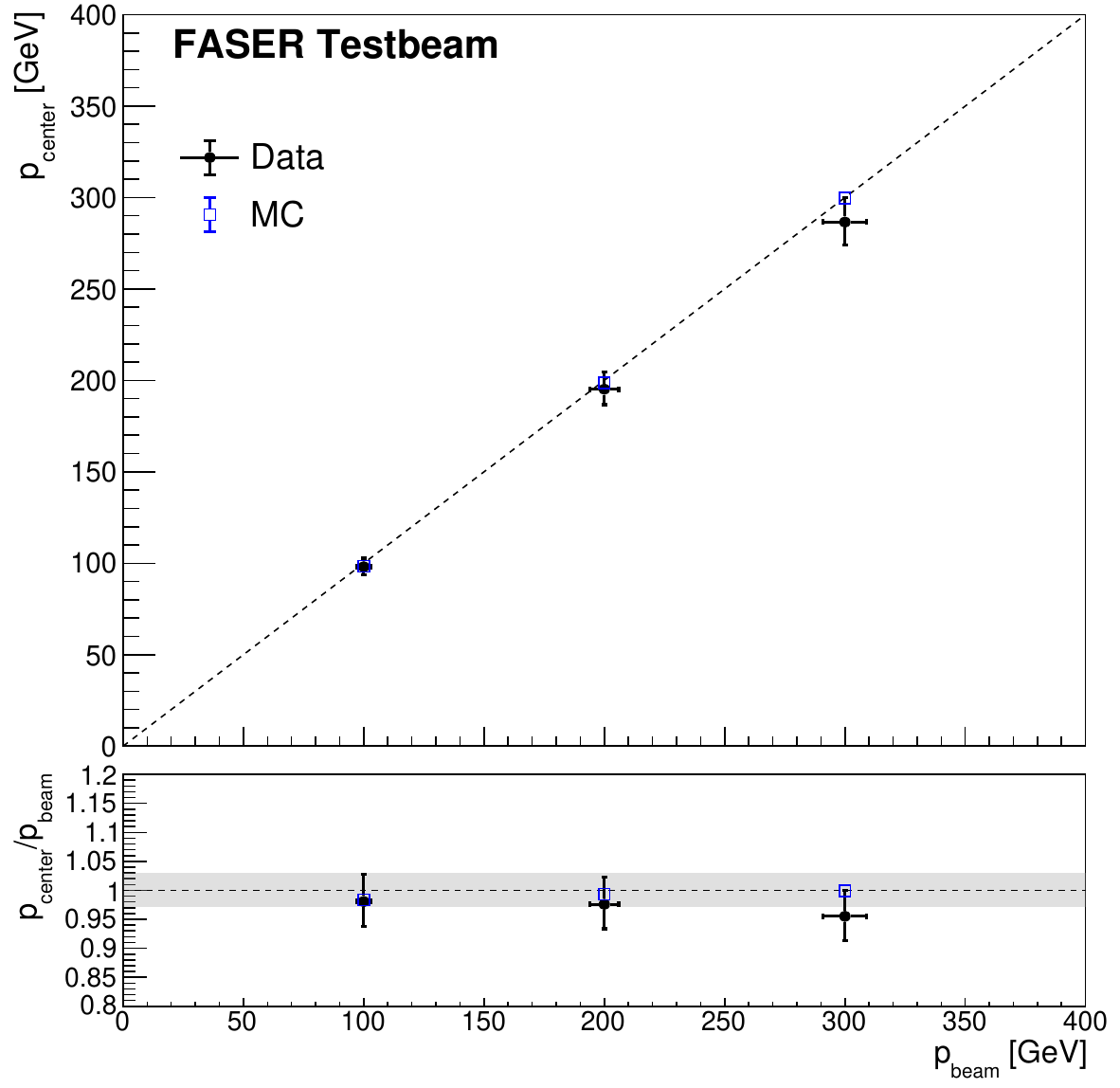}
    \end{minipage}
    \begin{minipage}{0.48\textwidth}
        \centering
        \includegraphics[width=\textwidth]{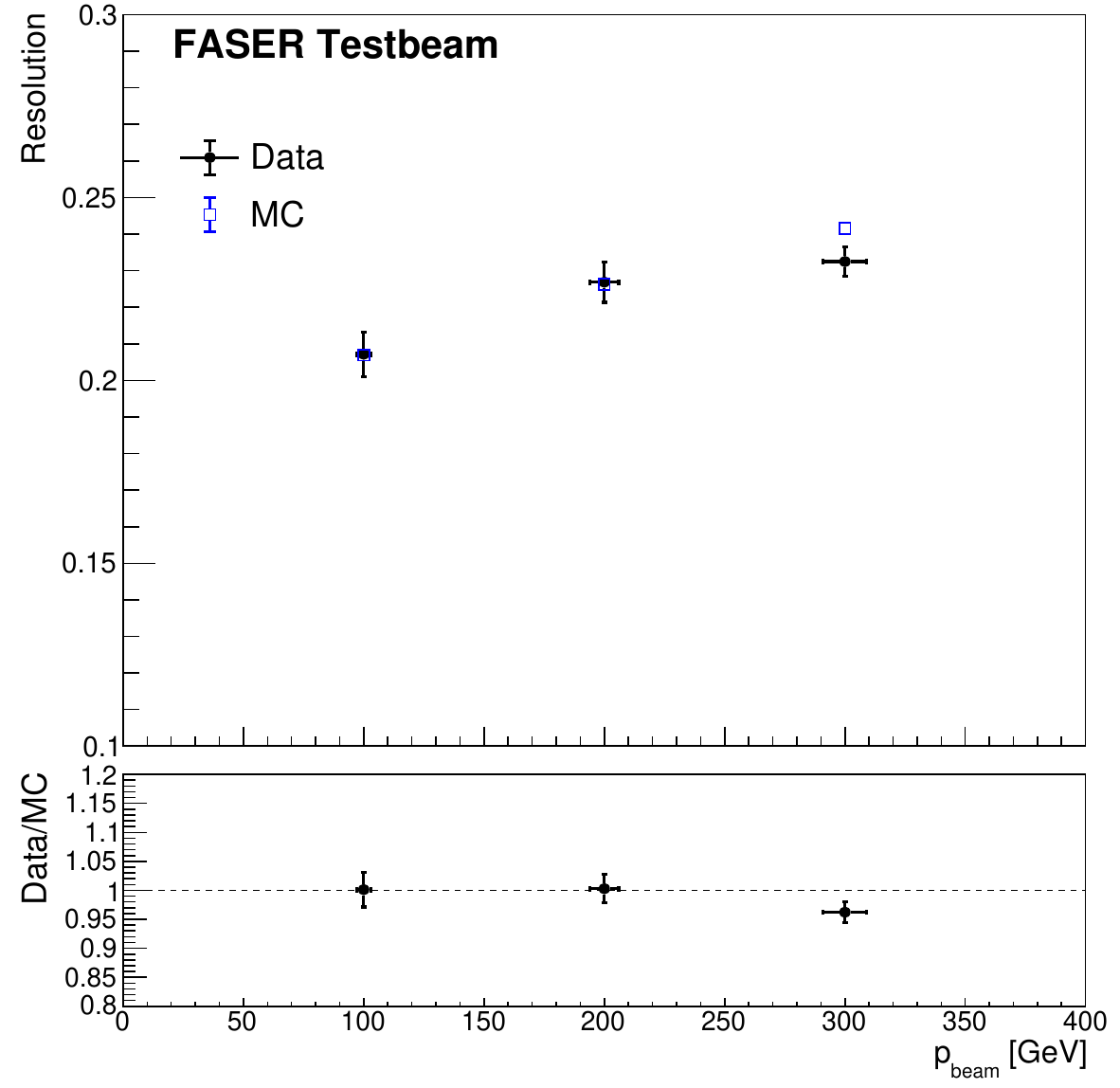}
    \end{minipage}
    \caption{Comparison between the test beam data and the MC simulation. Black circles represent the data, and blue open squares represent the MC. The data were assigned a 3\% uncertainty on $p_{\mathrm{beam}}$. Top left: $p_{\mathrm{center}}$ as a function of $p_{\mathrm{beam}}$. Top right: Resolution as a function of $p_{\mathrm{beam}}$. Bottom left: Ratio of $p_{\mathrm{center}}$ for the data and MC to $p_{\mathrm{beam}}$. Error bars include only the uncertainty on $p_{\mathrm{center}}$, while the gray band indicates the 3\% beam momentum uncertainty propagated to the ratio. Bottom right: Ratio of the data to MC results for the resolution. For the 300 GeV case, the data exhibit a smaller resolution value than the MC, which is attributed to an underestimation of $p_{\mathrm{center}}$.}
    \label{fig:RatioDataMC}
\end{figure}

Momentum measurements were performed using muon tracks penetrating all 100 plates in each of the two areas.
The measurement condition was set to $n_{\mathrm{cell}}^{\mathrm{max}}$ = 24. \Cref{fig:MomentumMeasurementResult} shows the reconstructed momentum distributions (top) and the inverse reconstructed momentum distributions (bottom) for 100, 200 and 300 GeV muons from left to right. The black points represent the data, while the blue histogram represents the MC, which is the same sample used in \cref{sec:Evaluation}, except that the position smearing was changed from 0.3 to \SI{0.24}{\micro\metre}.
This result was obtained using tracks from both sub-areas, excluding tracks that overlap between areas A and B in the distributions shown in \cref{fig:MomentumMeasurementResult}.
As shown in \cref{fig:MomentumMeasurementResult}, the $p_{\mathrm{rec}}$ and $1/p_{\mathrm{rec}}$ distributions of data and MC are in good agreement.
The low-momentum peak observed in the 300 GeV data was found to originate from a background beam with a different angular spread. 
Therefore, for the 300 GeV beam, the MC distributions were normalized to the data using only events with reconstructed momenta above 100 GeV, while for the 100 and 200 GeV beams they were simply normalized to the total number of tracks in the data.
The values of $p_{\mathrm{center}}$ and measurement resolutions of $p_{\mathrm{rec}}$ are summarized in \cref{tab:ComparePcenterDataToMC}. 
The resolutions are derived from a Gaussian fit to the $1/p_{\mathrm{rec}}$ distribution shown in \cref{fig:MomentumMeasurementResult} (bottom), with the fitting example illustrated in \cref{fig:MomentumEvaluationExample} (right).
For the data, the uncertainty on $p_{\mathrm{center}}$ includes not only the measurement resolution but also the thickness uncertainty. 
The thickness uncertainty is evaluated from the difference between the reconstructed and measured average thicknesses.
The average thickness reconstructed from data is \SI{1467}{\micro\metre}, while the measured average thickness of the tungsten plates and emulsion films is \SI{1427}{\micro\metre}.
This difference of \SI{40}{\micro\metre} is taken as the thickness uncertainty.
Assuming that this difference originates from only the tungsten thickness, error propagation using \cref{eq:splanerms} gives an uncertainty of 4.5\% on $p_{\mathrm{center}}$.
Considering all sources of uncertainty, the total uncertainty on $p_{\mathrm{center}}$ is evaluated to be 4.5\% across all momenta.
The values of $p_{\mathrm{center}}$ obtained from the two different sub-areas agreed within this uncertainty.
In \cref{eq:splanerms}, the plate thickness for each cell is assumed to be constant.
In practice, however, the thickness can vary due to non-uniformities of the tungsten plates and alignment uncertainties.
The RMS/mean of the reconstructed thickness in data is measured to be 1.5\%.
The impact of such thickness variations on both $p_{\mathrm{center}}$ and the resolution was evaluated and found to be negligible compared to the statistical uncertainties in the data.
\Cref{fig:RatioDataMC} shows $p_{\mathrm{center}}$ (left) and the resolution (right) for data and MC as functions of $p_{\mathrm{beam}}$. 
The muon beam at the SPS-H8 beamline is known to have a momentum uncertainty of 2\% and spread of about 3\%. Therefore, a 3\% uncertainty was assigned to $p_{\mathrm{beam}}$.
In the left panel, $p_{\mathrm{center}}$ in data is consistent with $p_{\mathrm{center}}$ in MC and with $p_{\mathrm{beam}}$.
The right panel shows a measurement resolution of 20–23\% in the momentum range of 100–300 GeV.
Based on these results, the method is validated up to 300 GeV using the test beam data.


\section{Momentum measurement using the \FASERnu detector with the LHC beam}
\label{sec:UsingF222}
As a first probe of the momentum measurement method above 300 GeV, a subset of background muon data recorded by the \FASERnu detector between July 26th and September 13th, 2022, corresponding to an integrated luminosity of 9.5 fb$^{-1}$ of $pp$ collisions at a center-of-mass energy of 13.6 TeV, is examined.
The analyzed data consist of small regions of \SI{6.7}{\milli\metre} $\times$ \SI{7.6}{\milli\metre} in area, with tracks selected to penetrate more than 100 tungsten plates.

\begin{figure}[b]
    \centering
    \begin{minipage}{0.43\textwidth}
        \centering
        \includegraphics[width=\textwidth]{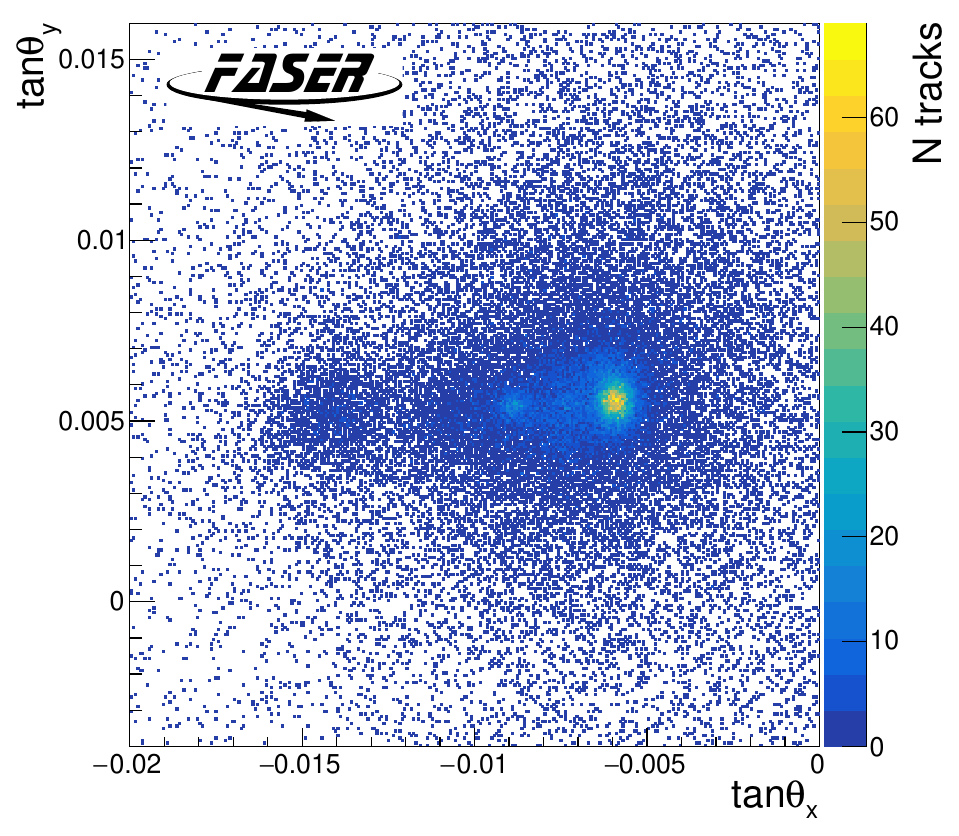}
    \end{minipage}    
    \begin{minipage}{0.43\textwidth}
        \centering
        \includegraphics[width=\textwidth]{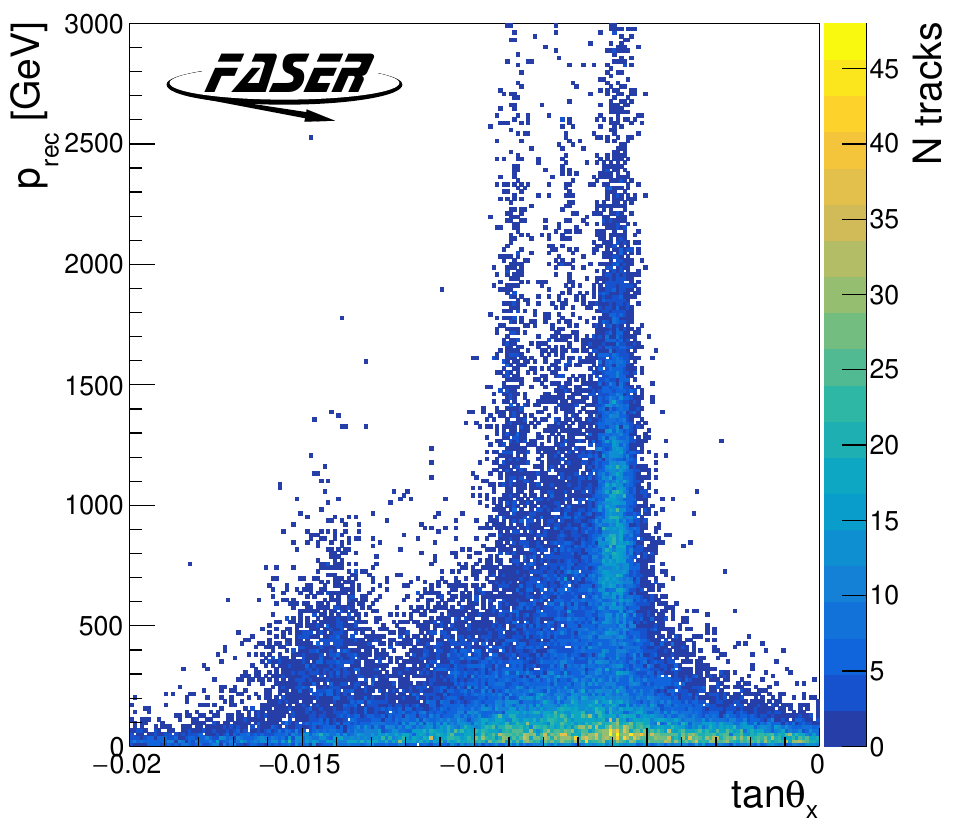}
    \end{minipage}
    \caption{Left: Two dimensional distribution of track angles. Right: Reconstructed momentum as a function of $\tan\theta_{x}$, with a selection on $\tan\theta_{y}$ applied in the range $-0.004 < \tan\theta_{y} < 0.016$.}
    \label{fig:TrackAngleDistributionAndPrecTxWithF222}
\end{figure}

Figure \ref{fig:TrackAngleDistributionAndPrecTxWithF222} (left) shows the two-dimensional distribution of the track angles of background muons.
The track angle was reconstructed from the position information of base-tracks within the first ten films, and three distinct peaks are clearly visible.
Since the \FASERnu box is not precisely aligned with the incoming neutrino beam direction, the $\tan\theta_x$ and $\tan\theta_y$ distributions do not have its main peak at zero.
The largest peak on the right is fitted with a Gaussian distribution and has a width of approximately \SI{420}{\micro rad}.
Assuming that this angular spread arises from multiple Coulomb scattering in the 100 m of rock located upstream of the \FASERnu detector, and using \cref{eq:thetaRMS}, we find that the observed angular spread can be attributed to muons with a momentum of $1290^{+150}_{-130}$ GeV.
A 10\% uncertainty is assumed for both the rock thickness and the radiation length. 
\Cref{fig:TrackAngleDistributionAndPrecTxWithF222} (right) shows the $p_{\mathrm{rec}}$ distributions as functions of the track angles, with $\tan\theta_x$.
The three angular peaks correspond to different momentum peaks, and the strongest one is centered around 1 TeV, with a 68\% interval of [450, 1620] GeV.
This is consistent with the estimation from the angular distribution and provides an indication that the method reconstructs muons with TeV-scale momenta.
In future analyses, by matching muon tracks between the \FASERnu detector and the FASER magnetic spectrometer located downstream of the \FASERnu detector for both neutrino-induced or background muon events, the momentum measurement results will be cross-validated.

\section{Conclusions}
\label{sec:Conclusions}

Momentum measurement is essential for the neutrino kinematic analysis in the FASER experiment. 
The momentum of charged particles can be determined from their deflections due to multiple Coulomb scattering (MCS).
Taking advantages of the high spatial resolution and long tracking length of the \FASERnu detector, the coordinate method based on MCS is applied to measure charged particle momenta from a few GeV to a few TeV.
The performance was evaluated using Monte Carlo simulations, and $n_{\mathrm{cell}}^{\mathrm{max}}$ = 24 showed the best performance using 100 plates with typical detector resolutions.
To validate this method, a test beam was performed at the H8 beamline of the CERN's SPS, and 100, 200 and 300 GeV $\mu^{-}$ beams were exposed to a detector comprising 100 plates.
The measured central momenta are consistent between data and MC within uncertainties. 
A momentum resolution of 20–23\% is achieved in the 100–300 GeV range.
Using background muons recorded in the \FASERnu detector in the LHC beam, the momentum measurement method was applied as a first probe of muons with TeV-scale momenta.
The momentum inferred from the angular spread and the reconstructed momentum distribution are consistent. These results motivate future cross-validation using matched tracks between the \FASERnu detector and the FASER spectrometer.


\section*{Acknowledgments}
\label{sec:Acknowledgments}

We thank CERN for the excellent performance of the LHC and the technical and administrative staff members at all FASER institutions for their contributions to the success of the FASER experiment. We thank the CERN EN-HE group for their help with the installation/removal of the FASER$\nu$ detector, and the CERN EP department for the refurbishment of the CERN dark room, used for the preparation and development of the FASER$\nu$ emulsion. We thank Saya Yamamoto for supporting the preparation of emulsion films. We thank the CERN SPS team for the efficient operation of the beam line and for the excellent beam conditions during the test-beam campaign.
We thank Maarten Van Dijk for useful discussions on the SPS test beam operation and for providing valuable information on the beam characteristics.

This work was supported in part by Heising-Simons Foundation Grant Nos.~2018-1135, 2019-1179, and 2020-1840, Simons Foundation Grant No.~623683, U.S. National Science Foundation Grant Nos.~PHY-2111427, PHY-2110929, and PHY-2110648, JSPS KAKENHI Grant Nos.~19H01909, 22H01233, 20K23373, 23H00103, 20H01919, 21H00082 and 25KJ0719, the joint research program of the Institute of Materials and Systems for Sustainability, ERC Consolidator Grant No.~101002690, BMBF Grant No.~05H20PDRC1, DFG EXC 2121 Quantum Universe Grant No.~390833306, Royal Society Grant No.~URF$\backslash$R1$\backslash$201519, UK Science and Technology Funding Councils Grant No.~ST/ T505870/1, the National Natural Science Foundation of China, Tsinghua University Initiative Scientific Research Program, and the Swiss National Science Foundation.

\bibliographystyle{utphys}
\bibliography{references}

\end{document}